\documentclass[twocolumn]{aastex631}

\newcommand\Msun{\; {\rm M}_{\odot}}
\newcommand\kms{\; {\rm km}\;{\rm s}^{-1}}
\newcommand\pc{\;{\rm pc}}
\newcommand\kpc{\;{\rm kpc}}

\newcommand\Myr{\;{\rm Myr}}
\newcommand\Gyr{\;{\rm Gyr}}

\newcommand\QTmin{{Q_{T,\text{min}}}}
\defcitealias{jnk23}{JK23}

\usepackage{amsmath}
\usepackage{graphicx}
\usepackage[nolist]{acronym}
\usepackage[nameinlink,capitalise,noabbrev]{cleveref}

\shorttitle{Effects of Halo Spin on Bar Formation in Disk Galaxies}
\shortauthors{JANG \& KIM}

\begin{document}

\title{Effects of Halo Spin on the Formation and Evolution of Bars in Disk Galaxies}

\correspondingauthor{Woong-Tae Kim}
\email{unitree@snu.ac.kr}

\author[0000-0002-7202-4373]{Dajeong Jang} 
\affiliation{Department of Physics $\&$ Astronomy, Seoul National University, Seoul 08826, Republic of Korea}

\author[0000-0003-4625-229X]{Woong-Tae Kim}
\affiliation{Department of Physics $\&$ Astronomy, Seoul National University, Seoul 08826, Republic of Korea}
\affiliation{SNU Astronomy Research Center, Seoul National University, Seoul 08826, Republic of Korea}

\begin{acronym}
    \acro{BPS}{boxy peanut-shaped}
\end{acronym}

\begin{abstract}  
The spin of dark halos has been shown to significantly affect bar formation and evolution in disk galaxies. To understand the physical role of the halo spin on bar formation, we run $N$-body simulations of isolated, Milky Way-sized galaxies by varying the halo spin parameter in the range $-0.16 \leq \lambda \leq 0.16$ and the bulge mass. We find that our adopted halo \emph{alone} is subject to swing amplification of an $m=2$ non-axisymmetric mode rotating in the same sense as the halo, which assists or inhibits the bar formation in a disk depending on its sense of rotation. The $m=2$ mode in the disk, growing via swing amplification, interacts constructively (destructively) with the $m=2$ mode in the prograde (retrograde) halo, promoting (delaying) bar formation. A bar grows by losing its angular momentum primarily to a halo.  Since the halo particles inside (outside) the corotation resonance with the bar can emit (absorb) angular momentum to (from) the bar, the bar pattern speed decays slower for larger $\lambda>0$, while it decreases relatively fast almost independent of $\lambda\leq0$.  Models with a strong bar develop a boxy peanut-shaped bulge. In models without a bulge, this occurs rapidly via buckling instability, while the bars with a bulge thicken gradually without undergoing buckling instability. Among the models considered in the present work, the bar in the $\lambda = 0.06$ model with a bulge of 10\% of the disk mass best describes the Milky Way in terms of the bar length and pattern speed. 
\end{abstract}


\section{Introduction} \label{sec:intro}
 
Nearly $2/3$ of disk galaxies in the local universe possess a weak or strong bar in optical and near-infrared images \citep{de63, whyte02,lauri04,marijogee07,menendez07,agu09,mendez12,buta15,diaz16,diaz19}. Bars are one of the main drivers of the dynamical and secular evolution of disk galaxies, responsible for stellar migration \citep{di13,kawata17,halle18,iles24}, angular momentum transport from disk to halo \citep{sn13,peter16,col18,col21}, and formation of pseudo bulges \citep{comb81,comb90,ath05,deba06,gadotti11}. Despite the importance of bars, however, there has been no clear answer as to why some galaxies are barred or non-barred and what controls the physical properties of bars.

Many numerical studies have attempted to find one-parameter conditions for bar formation using the initial galaxy properties such as the ratio of the kinetic to gravitational potential energy and the ratio of the bulge to disk mean density \citep{onp73,efsta82, kd18, se18}. While the proposed conditions explain the bar formation in their own adopted models, they fail to be applicable to diverse galaxy models. The most likely reasons for this may be that the proposed conditions are based on a limited set of galaxy models by fixing either bulge or halo parameters and that the one-parameter conditions cannot capture the complicated processes of bar formation. Recently, \citet[hereafter \citetalias{jnk23}]{jnk23}  proposed the two-parameter condition 
\begin{equation}\label{e:cri}
    \left(\frac{\QTmin}{1.2}\right)^2 + \left(\frac{\text{CMC}}{0.05}\right)^2\lesssim 1
\end{equation}
for bar formation. In \cref{e:cri},  $\QTmin$ is the minimum value of the Toomre stability parameter of the disk and CMC denotes the central mass concentration. This condition is physically motivated: bar formation requires several cycles of swing amplification and feedback loops \citep{sell80,toomre81,bnt08}, and perturbations grow more easily in a disk with smaller $\QTmin$ and CMC. 
\cref{e:cri} implies that the presence of a (massive) halo tends to suppress bar formation by increasing both $\QTmin$ and CMC, which is qualitatively consistent with the established result that the density of a halo has a negative influence on the bar formation \citep{onp73,hohl76,efsta82,deba00}.

While \cref{e:cri} accounts for bar formation in diverse galaxy models with various degrees of the central concentration of halo and bulge, it is limited to models with a non-rotating halo. Cosmological simulations of galaxy formation found that dark halos acquire angular momentum from tidal torque \citep[e.g.,][]{hoy51,fal80,whi84} and satellite accretion \citep[e.g.,][]{vit02,dilla23}. 
As a dimensionless measure of halo spin, \citet{bull01} introduced
\begin{equation}\label{eq:lambda}
\lambda \equiv \frac{L_\mathrm{vir}}{\sqrt{2}M_\mathrm{vir}r_\mathrm {vir}V_c},
\end{equation}
where $L_{\rm vir}$ and $M_\mathrm{vir}$ are the angular momentum and mass within the virial radius $r_\mathrm{vir}$, respectively, and $V_c=(GM_\mathrm{vir}/r_\mathrm{vir})^{1/2}$ is the circular velocity at $r_\mathrm{vir}$.\footnote{\Cref{eq:lambda} was derived from  $\lambda_P=L|E|^{1/2}/(GM^{5/2})$, where $L$, $E$, and $M$ refer to the total angular momentum, energy, and mass of a spherical system within radius $r$, respectively (\citealt{peebles69,peebles71}; see also \citealt{ans23}). At the virial radius where $E=-GM_\mathrm{vir}/(2r_\mathrm{vir})$, $\lambda$ in \cref{eq:lambda} becomes identical to $\lambda_P$.} Note that positive (negative) $\lambda$ corresponds to a spinning halo in the same (opposite) sense as the disk rotation. Cosmological simulations showed that $\lambda$ follows a lognormal distribution, with the mean value $\lambda \simeq 0.03$--$0.04$ \citep{bull01,hetz06,bett07,jia19}.
The Milky Way halo is also thought to spin with $\lambda=0.061$ \citep{obr22}.

$N$-body simulations for bar formation with a spinning halo commonly found that a stellar disk embedded in a prograde halo with larger $\lambda\; (>0)$ develops a bar faster \citep{sn13,long14, col18,col21,kns22,li24}, while bar formation in a disk under a retrograde halo with smaller $\lambda\;(<0)$ is progressively delayed \citep{col19b}. To explain this, \citet{sn13} invoked the condition $t_{\rm OP} \equiv T_\mathrm{rot}/|W|>0.14$ of \citet{onp73} for bar formation, where $T_\mathrm{rot}$ and $W$ refer to the rotational and gravitational potential energies of a galaxy, respectively. They argued that the halo spin increases $T_\mathrm{rot}$ to form a bar faster. Since \citet{onp73} employed models with a non-rotating, fixed halo, however, it is questionable whether $T_\mathrm{rot}$ should include the disk rotation only or the halo rotation as well.  Even if the halo rotation can be included, $t_{\rm OP}$ is independent of the sense of halo rotation, as \citet{sn13} noticed, which is not compatible with the numerical results: a bar forms earlier in a model with $\lambda\;(>0)$ than in the $-\lambda$ counterpart. This makes $t_{\rm OP}$ inadequate for explaining the effect of the halo spin on bar formation.  

Once a bar forms, the angular momentum transfer between it and its surrounding halo affects the bar growth significantly. When $\lambda=0$, a bar grows by losing its angular momentum to a halo  (\citealt{ath02,mar06,kd18,se18}; \citetalias{jnk23}). The angular momentum transfer appears more complicated when the halo spin is considered. \citet{long14} found that the inner part of a prograde halo loses its angular momentum to a bar, while the outer part acquires it.
\cite{col19a} suggested that the angular momentum transfer by the resonances is more active under a halo with higher $\lambda>0$.
\citet{col19b} and \citet{col21} proposed that orbital reversals of halo particles are a crucial factor of angular momentum transfer between disk and halo with $\lambda<0$. 
\cite{kns22} found that it is the spin of the inner halo (with $R < 30\kpc$) rather than the total halo spin that is important for the bar formation and its ensuing evolution.

In this paper, we use $N$-body simulations to investigate the effects of halo spin and the presence of a classical bulge on the bar formation and evolution. Our primary purpose is to
understand how the halo spin physically affects the bar formation. For this, we consider Milky Way-sized galaxy models with differing bulge mass and halo spin. To delineate the effect of halo spin, we consider a wide range of the spin parameter $-0.16 \leq \lambda \leq 0.16$, and compare the results from halo-only models with $\lambda=\pm0.16$ or 0.  We monitor the temporal changes of the bar strength and pattern speed, study the vertical buckling of bars in our simulations, and analyze their dependencies on the spin parameter. 

This paper is structured as follows. In \autoref{sec:modelnmethod}, we describe our galaxy models and numerical methods. In  \autoref{sec:halo},  we run simulations without a disk and bulge to study the gravitational susceptibility of our adopted halos.
In \autoref{sec:results}, we present the results of full galaxy models, focusing on the dependence on $\lambda$ of the bar formation, evolution, pattern speed, and angular momentum transfer. We also present the results for vertical buckling instability of our models and its outcomes.  
In \autoref{sec:discussion}, we discuss our results in comparison with the previous results and constrain the model parameters for the Milky Way. 
Finally, we conclude our work in \autoref{sec:summary}.

\section{Galaxy Model and Method} \label{sec:modelnmethod}

\begin{figure}[t]
\centering
\plotone{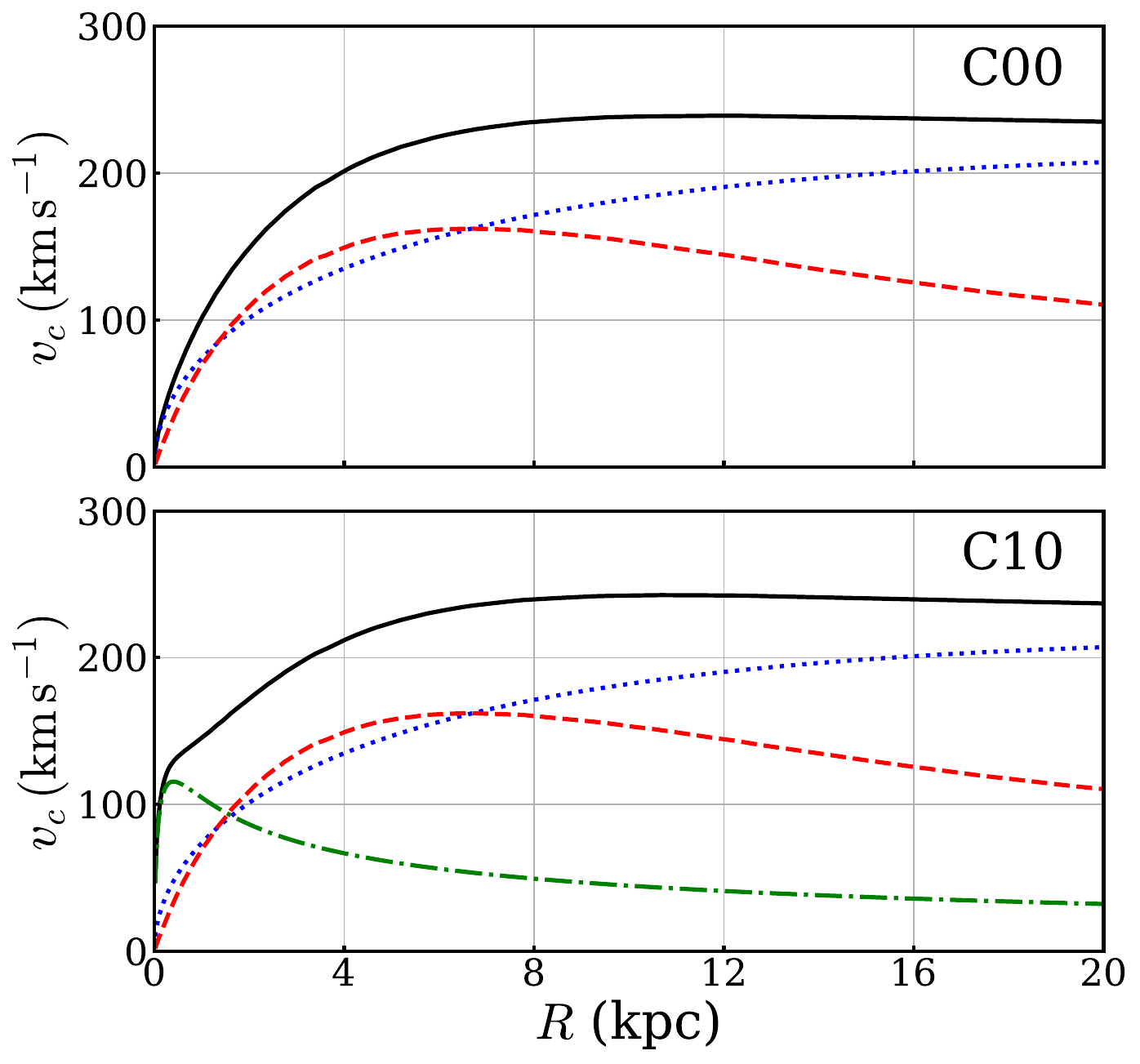}
\caption{Radial distributions of the total circular velocity $v_c$ (black solid)  as well as the contributions of a halo (blue dotted), a disk (red dashed), and a bulge (green dot-dashed). The \texttt{C00} and \texttt{C10} series in the upper and lower panels refer to the models without and with a classical bulge, respectively.
\label{fig:vrot}}
\end{figure}

To study the effects of halo spin on bar formation and evolution, we consider isolated galaxy models whose mass and size are similar to those of the Milky Way. Our galaxy models consist of a dark matter halo, a stellar disk, and a central supermassive black hole. Some models do not have a classical bulge (\texttt{C00} series), but we also consider models with a bulge (\texttt{C10} series). 

The stellar disk has the conventional exponential-secant hyperbolic density distribution
\begin{equation}
\rho_d(R,z) = \frac{M_d}{4\pi z_dR_d^2} \exp\left( -\frac{R}{R_d}\right)
{\rm sech}^2\left( \frac{z}{z_d}\right),
\end{equation}
where $R$ is the cylindrical radius, $z$ is the vertical height, $R_d$ is the disk scale radius, $z_d$ is the disk scale height, and $M_d$ is the total disk mass.
We take $R_d=3\kpc$, $z_d=0.3\kpc$, and $M_d=5\times 10^{10}\Msun$, similar to the Milky Way (\citealt{bg16}, \citealt{helmi20}).

\begin{deluxetable}{cccr}
\tablecaption{Model  parameters \label{tbl:model}}
\tablenum{1}
\tablehead{
\colhead{Name} & 
\colhead{\hspace{.2cm}$M_b/M_d$}\hspace{.2cm} & 
\colhead{\hspace{.2cm}$\lambda$}\hspace{.2cm} &
\colhead{\hspace{.1cm}$f_p$}} 
\startdata
\texttt{C00P16}   & \nodata        & $0.16$   &  1.00   \\
\texttt{C00P10}   & \nodata        & $0.10$   &  0.80  \\
\texttt{C00P06}   & \nodata        & $0.06$   &  0.67  \\
\texttt{C00P00}   & \nodata        & $0.00$  &  0.50  \\
\texttt{C00R06}   & \nodata        & $-0.06$  &   0.32   \\
\texttt{C00R10}   & \nodata        & $-0.10$   &  0.20   \\
\texttt{C00R16}   & \nodata        & $-0.16$  &  0.00   \\
\hline
\texttt{C10P16}   & 0.1         & $0.16$  &  1.00    \\
\texttt{C10P10}   & 0.1        & $0.10$   &  0.80   \\
\texttt{C10P06}   & 0.1         & $0.06$  &   0.67   \\
\texttt{C10P00}   & 0.1         & $0.00$  &   0.50 \\
\texttt{C10R06}   & 0.1         & $-0.06$  &   0.32   \\
\texttt{C10R10}   & 0.1        & $-0.10$   &  0.20   \\
\texttt{C10R16}   & 0.1         & $-0.16$  &   0.00 
\enddata 
\end{deluxetable}

Both the halo and classical bulge have a density distribution that follows  \cite{hernquist90} profile
\begin{equation}
\rho(r) = \frac{M}{2\pi}\frac{a}{r(r+a)^3},
\end{equation}
where $r=(R^2+z^2)^{1/2}$ is the spherical radius, and $M$ and $a$ indicate the mass and the scale radius of each component, respectively.
For the halo, we fix its mass and scale radius to $M_{h}=1.3\times 10^{12}\Msun=26M_d$ and $a_h=30\kpc$, similar to the Milky Way (e.g., \citealt{bg16}, \citealt{helmi20}). 
For the bulge, we set its scale radius to $a_b=0.4\kpc$ and mass to $M_b = 0.1 M_d$ (e.g., \citealt{shen10}, \citealt{bg16}, \citealt{helmi20}). 
We place a supermassive black hole with mass $M_\mathrm{BH}=4 \times 10^{6}\Msun$ at the galaxy center  (\citealt{ghez08,gill09a,gill09b}). 
\autoref{fig:vrot} plots the total circular velocity $v_c$ calculated from the gravitational potential and the contribution of each component for the models in the \texttt{C00} and \texttt{C10} series. Due to the bulge, $v_c$ in the \texttt{C10} series increases rapidly at small $R$ compared to that in the \texttt{C00} series.

To assign the spin parameter $\lambda$ to the dark halo, we follow the method of \citet{long14} and \citet{col18,col19b}. That is, we randomly select a fraction of the halo particles and change the directions of their tangential velocities, while keeping their amplitudes and radial velocities intact. While \citet{kns22} explored the models in which the fraction $f_p$ of the halo particles on prograde orbits with respect to the disk rotation varies with radius, we keep $f_p$ spatially constant for simplicity. It turns out that our halo has $\lambda=0.16$ when $f_p=1$. The prograde fraction $f_p$ decreases as $\lambda$ decreases, leading to $\lambda=0$ for $f_p=0.5$, and $\lambda=-0.16$ for $f_p=0$. This simple method of the velocity reversals does not change the Boltzmann equation and thus makes the density profile of the halo unchanged by varying $\lambda$ \citep{lyn60,wein85,long14,col18,col19a}.
In our halo model, the maximum and minimum value of $\lambda$ is $\pm0.16$, corresponding to either all prograde or all retrograde particles.

\autoref{tbl:model} lists the names, bulge-to-disk mass ratio, spin parameter $\lambda$, and prograde fraction $f_p$ of our models. The models in the \texttt{C00} series do not possess a bulge, while those in the \texttt{C10} series have a bulge whose mass is 10\% of the disk mass. The infixes \texttt{P} and \texttt{R} stand for a prograde and retrograde halo, respectively, and the numbers after the infixes denote $\lambda$ multiplied by 100. Note that the models \texttt{C00P00} and \texttt{C10P00} have a non-spinning halo, and are identical to models \texttt{C00} and \texttt{C10} presented in \citetalias{jnk23}, respectively. In the present work, we do not consider the rotation of a bulge: the effect of bulge rotation was recently studied by \citet{li24}. 

\begin{figure}[t]
\centering
\plotone{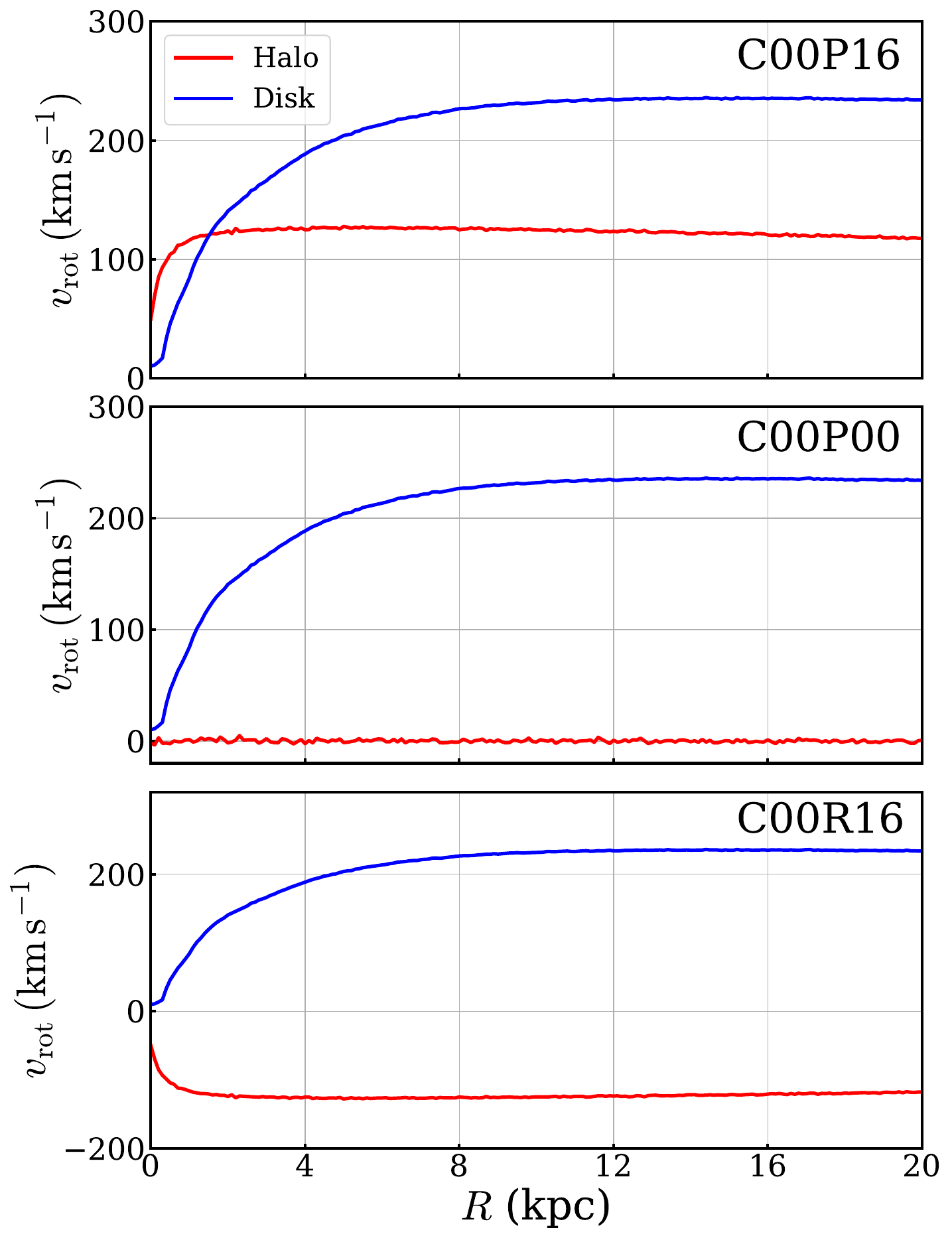}
\caption{Radial distributions of the halo and disk rotational velocity $v_\mathrm{rot}$ for \texttt{C00P16}, \texttt{C00P00}, and \texttt{C00R16}.
\label{fig:vrot_spin}}
\end{figure}

\autoref{fig:vrot_spin} compares the rotational velocities  $v_\mathrm{rot}=\langle  v_\phi \rangle$ for the disk and halo in models \texttt{C00P16}, \texttt{C00P00}, and \texttt{C00R16}.  Here, $v_\phi$ is the azimuthal velocities of the particles and the angle brackets denote the azimuthal and vertical averages. 
Note that  $v_\mathrm{rot}$ is slightly smaller than $v_c$ shown in \autoref{fig:vrot}, corresponding to asymmetric drift \citep[e.g.,][]{Oh08}.
In model \texttt{C00P00}, the rotational velocity of the halo is nearly zero. However, when all halo particles move in the same azimuthal direction, the rotational velocity of the halo particles is almost flat, amounting to $v_\mathrm{rot}\sim \pm 120\kms$ at $R\gtrsim 1\kpc$.  We will show in \autoref{sec:halo} that the halo rotation and velocity shear results in the creation of spirals, evolving into a weak bar, via swing amplification of density perturbations even in a halo-only system with $\lambda=\pm0.16$. 

To construct our galaxy models, we utilize the GALIC code \citep{yu14} which finds a desired equilibrium by adjusting the velocities of individual particles to the level that satisfies the collisionless Boltzmann equations.
The number of particles used is $N_d = 1.0 \times 10^6$, $N_b = 1.0 \times 10^5$, and $N_h = 2.6 \times 10^7$ for the disk, bulge, and halo, respectively. The mass of a single particle is set to $\mu=5\times10^4\Msun$, which is equal for all three components.

We evolve our galaxy models until $t=10\Gyr$ using a public version of the Gadget-4 code \citep{springel21}. 
We take the multipole expansion of order $p=4$ for fast force evaluation and a hierarchical time-integration scheme to reduce the computation time.
We fix the force accuracy parameter to $\alpha=3\times10^{-4}$ and the softening parameter to $0.05\kpc$, $0.01\kpc$, and $0.01\kpc$ for the halo, disk, and bulge particles, respectively, which we check conserves the total angular
momentum within $\sim0.1\%$.

\begin{figure} 
\centering
\plotone{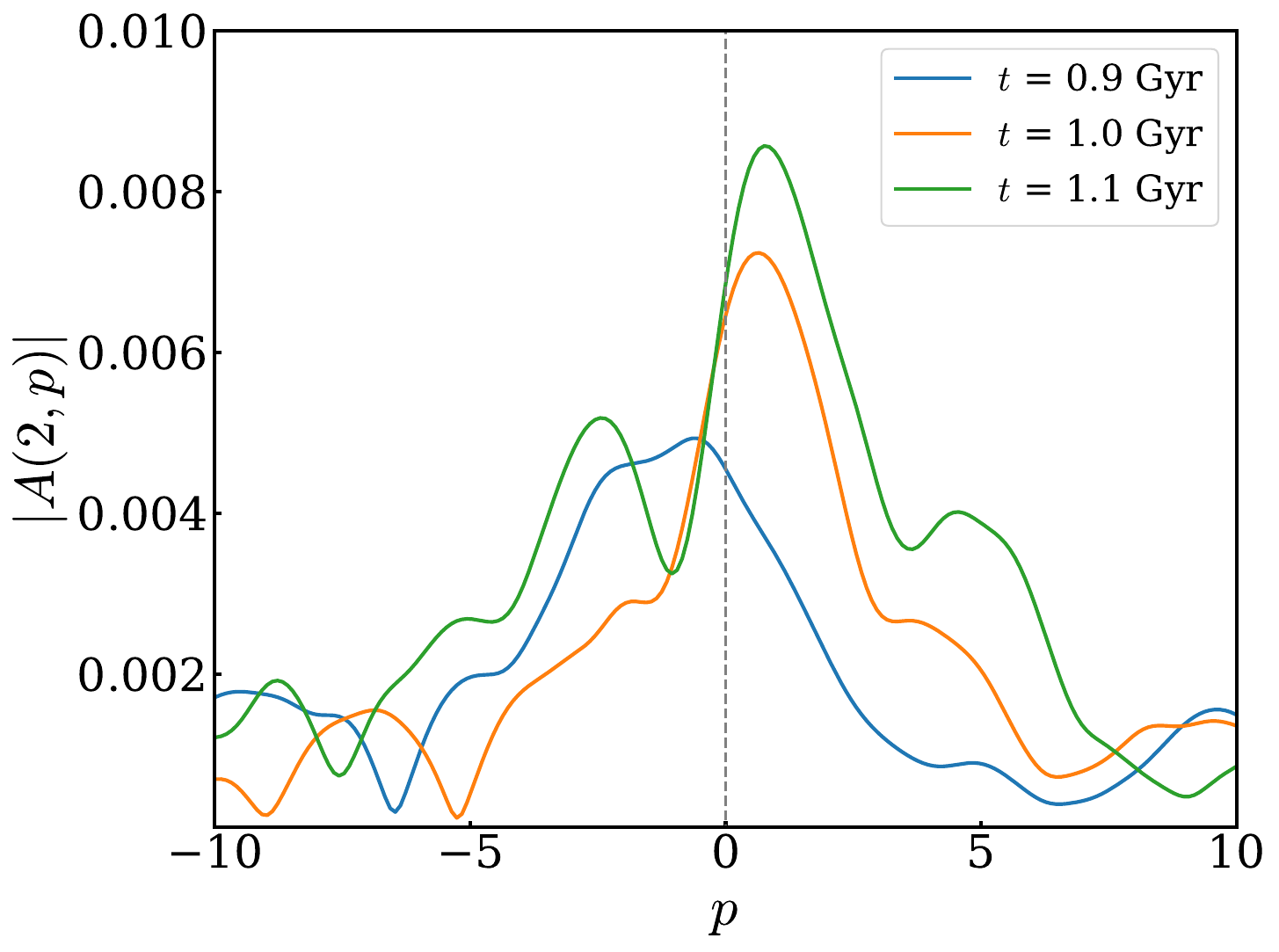}
\caption{Evolution of the Fourier coefficients with dimensionless wavenumber $p$, defined in \cref{e:logarm}, of the $m=2$ logarithmic arms in the regions with $R\leq 5\kpc$ and $|z|\leq 5\kpc$ of the halo-only model with $\lambda=0.16$. The dominant mode at $t=0.9, 1.0, 1.1\Gyr$ has $p=-0.6, 0.6, 0.8$, respectively. \label{fig:swing}}
\end{figure}

\begin{figure*} 
\centering
\plotone{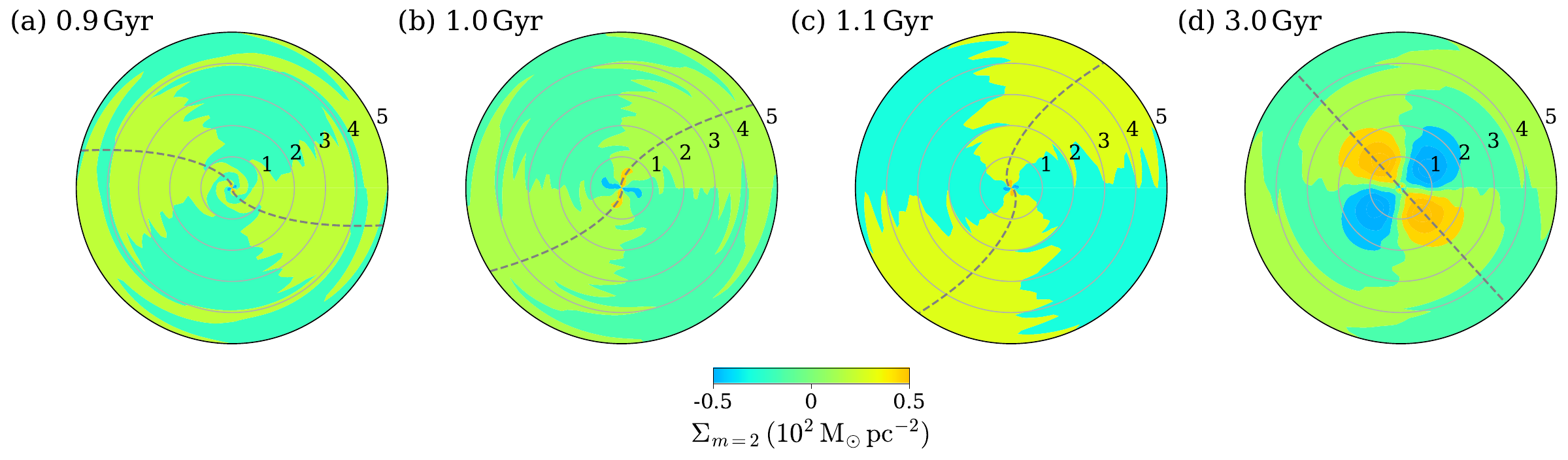}
\caption{Projected density distributions of the $m=2$ mode in the regions with $|z| \leq 5 \kpc$ and $R\leq 5\kpc$ for the halo-only model with $\lambda=0.16$
at (\emph{a}) $t = 0.9 \Gyr$, (\emph{b}) $t = 1.0 \Gyr$, 
(\emph{c}) $t = 1.1 \Gyr$, and (\emph{d}) $t = 3.0 \Gyr$. In each panel, 
the concentric rings indicate the radii from 1 to $5\kpc$. 
The dashed lines draw the crest of the dominant modes determined by \cref{e:logarm}. Since the halo is rotating in the counterclockwise direction, the dominant waves in (\emph{a}) are leading, while they are trailing in (\emph{b}) and (\emph{c}). The dominant mode is a bar in (\emph{d}). 
\label{fig:halo_spiral}}
\end{figure*}

\section{Halo-only Models} \label{sec:halo}

A rotating halo may be prone to forming weak, non-asymmetric structures via swing amplification, just as in a rotating disk. To explore this possibility, we run $N$-body simulations of halo-only models with $\lambda = 0.0$ (no spin), $0.16$ (prograde), and $-0.16$ (retrograde), constructed in \autoref{sec:modelnmethod}. 
By evolving these isolated halo models up to $t=2 \Gyr$, we confirm that the azimuthally-averaged, radial distributions of halo density, velocities, and velocity dispersions remain unchanged, indicating that the system is overall in dynamical equilibrium. At the same time, we also find that the rotating halo allows small-amplitude, non-axisymmetric perturbations to grow in regions with $R\leq 5\kpc$ and $|z|\leq 5\kpc$.

To quantify the strength of $m=2$ perturbations that grow in the halo, {we define the Fourier coefficients in $\ln R$ and $\phi$ as
\begin{equation}\label{e:logarm}
A(m,p) = \frac{1}{N}\sum_{j=1}^N 
\exp[i(m\phi_j + p \ln R_j)],
\end{equation}
where $N$ is the total number of particles and $(R_j, \phi_j)$ are the radial and azimuthal coordinates of the $j$-th particle in the regions with $R\leq 5\kpc$ and $|z|\leq 5\kpc$. In \cref{e:logarm}, the dimensionless radial wavenumber $p$ is related to the pitch angle $i=\tan^{-1}(m/p)$ of $m$-armed logarithmic spirals, and $p=0$ corresponds to a bar mode (e.g., \citealt{sell84,sell86,Oh08}).

\autoref{fig:swing} plots the temporal evolution of $|A(2, p)|$ of the $m=2$ logarithmic spiral waves in the halo-only model with $\lambda=0.16$ at $t=0.9$--$1.1\Gyr$. The perturbations are dominated by the mode with $p=-0.6$ at $t=0.9\Gyr$ and $p=0.8$ at $t=1.1\Gyr$. The corresponding projected distributions of the $m=2$ spiral waves, and the loci of the dominant modes are presented in \autoref{fig:halo_spiral}(\emph{a})--(\emph{c}). The waves grow as they swing from leading (negative $p$) to trailing (positive $p$) configurations.

\autoref{fig:halo_bar} plots the temporal changes in $|A(2, 0)|$,  the strength of a bar mode with $p=0$, in the halo-only models with $\lambda=0.16$  (red), $\lambda=0$  (black), and $\lambda=-0.16$ (blue), as well as its position angle 
\begin{equation}
  \psi(R)\equiv \frac{1}{2} \tan^{-1} \left[\frac{\sum_j \sin(2\phi_j)}{\sum_j\cos(2\phi_j)}\right],
\end{equation} 
measured at $R=2\kpc$ for the $\lambda=0.16$ model.  When the halo does not spin, the perturbations remain very weak with $|A(2, 0)|\lesssim 5 \times 10^{-3}$, primarily due to Poisson noise in
the particle distribution.\footnote{By running models with differing $N_h$, we have confirmed that $|A(2, 0)|\propto N_h^{-1/2}$ in the models with $\lambda=0$.} However, density perturbations in the rotating halos are swing-amplified to achieve $|A(2, 0)|\sim0.04$ at $t\sim2.4$--$2.8\Gyr$ in models with $\lambda=\pm0.16$. The small difference in $|A(2, 0)|$ between the prograde and retrograde halos is presumably caused by the amplitudes of the initial leading perturbations most susceptible to swing amplification.\footnote{Since the initial particle distribution is random, the power spectrum of the corresponding density perturbations is not symmetric in the radial wavenumber $p$.  The leading waves with negative $p$ in the prograde halo become trailing waves with positive $p$ in the retrograde halo.} 

\begin{figure} 
\centering
\plotone{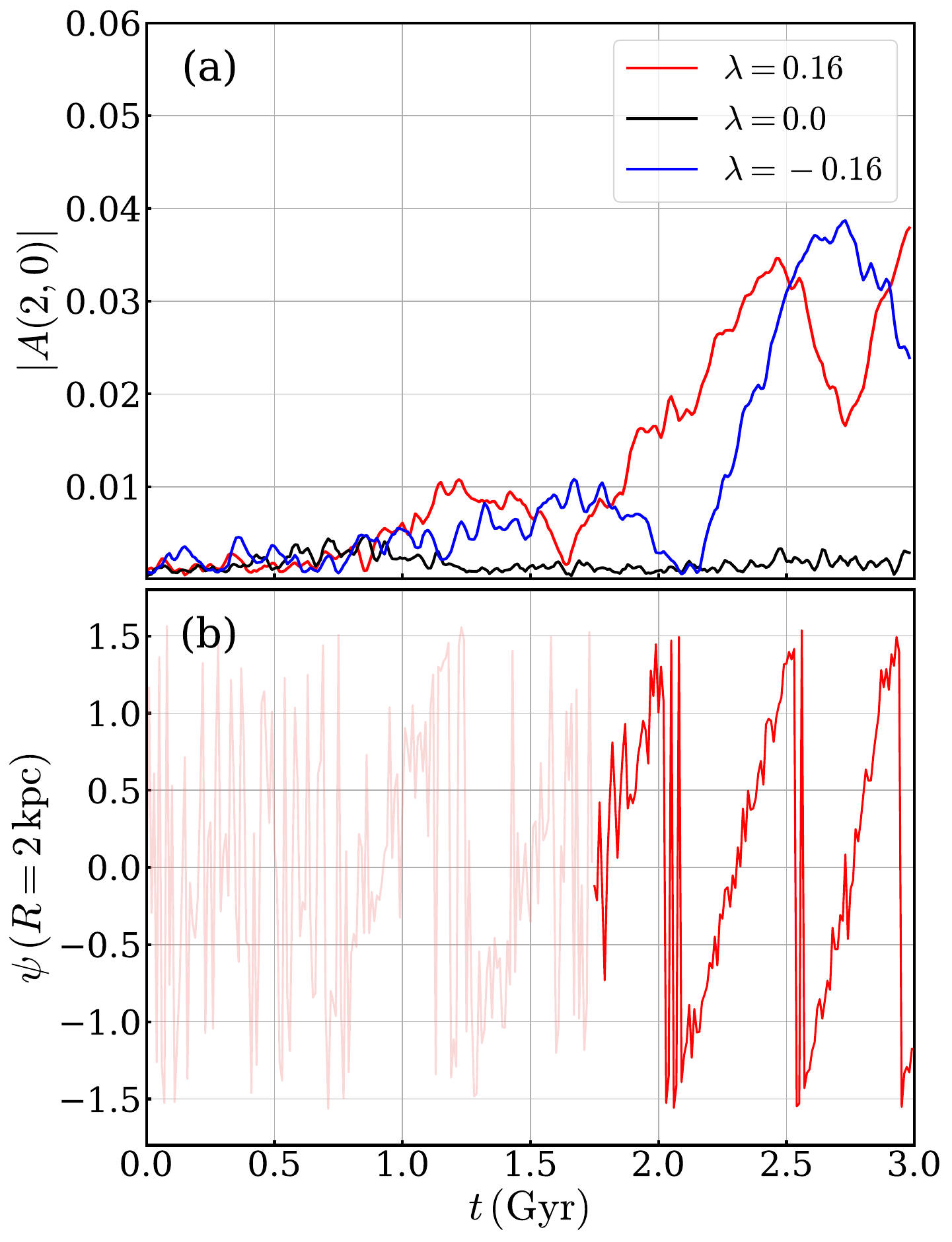}
\caption{ 
(\emph{a}) Temporal changes in the strength of the bar mode in the regions with $R\leq 5\kpc$ and $|z|\leq 3\kpc$ of the halo-only models with $\lambda=0.16$ (red), $\lambda  = 0.0$ (black),  and $\lambda=-0.16$ (blue). (\emph{b}) Temporal evolution of the bar position angle $\psi$ measured at $R=2\kpc$ in the $\lambda=0.16$ model.
\label{fig:halo_bar}}
\end{figure}

\autoref{fig:halo_spiral}(\emph{d}) shows that the spirals eventually grow into a weak bar. By using the cross-correlation of the halo surface densities at two different epochs in the annular regions with width $\Delta R = 0.1 \kpc$ at $R= 1 \kpc$ for $t=2$--$3\Gyr$ (see \citealt{seo19}, \citetalias{jnk23}), we find that the bar in the halo-only model with $\lambda=0.16$ has a pattern speed of $\Omega_h \sim 6 \kms \kpc^{-1}$, rotating in the counterclockwise direction. Note that $\Omega_h\approx d\psi/dt$  for $t=2$--$3\Gyr$ shown in \autoref{fig:halo_bar}(\emph{b}). 
Similarly, the perturbed density in the retrograde halo with $\lambda=-0.16$ grows first into trailing spirals and then to a bar with a pattern speed of $\Omega_h \sim -6 \kms \kpc^{-1}$. These indicate that a spinning halo \emph{alone} can form a weak bar that rotates in the same direction as the halo.

\begin{figure*} 
\centering
\plotone{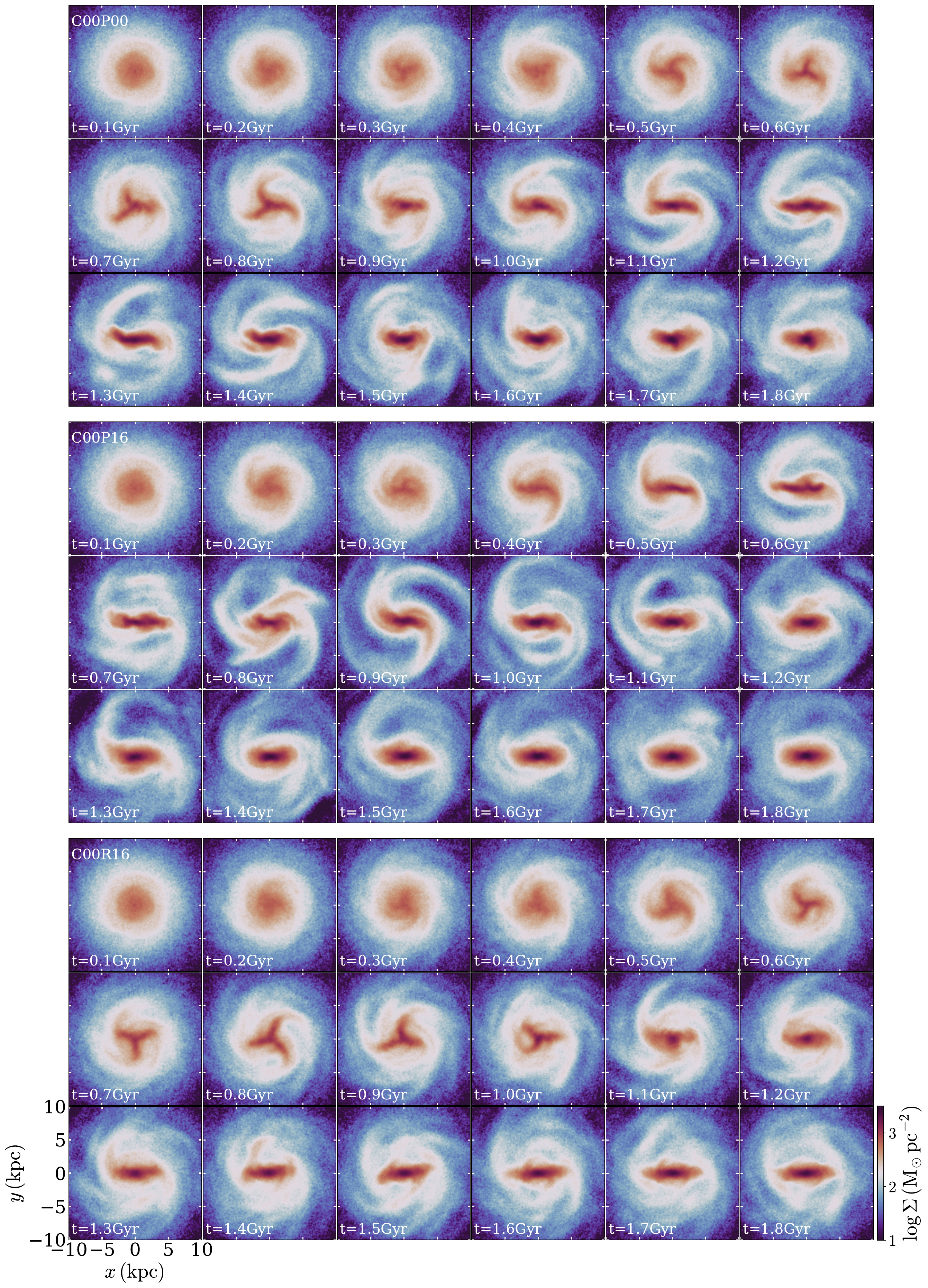}
\caption{Snapshots of the disk surface density at every $0.1\Gyr$ interval from $t=0.1$ to $1.8\Gyr$
in models \texttt{C00P00} with non-spinning halo (top), \texttt{C00P16} with a prograde halo (middle), and \texttt{C00R16} with a retrograde halo (bottom). \label{fig:snap}}
\end{figure*}

 Since swing amplification requires self-gravity, rotational shear, and epicyclic shaking \citep{toomre81}, it has been studied extensively in the context of galactic disks. However, it can also operate in a spherical system as long as the three agents mentioned above are present. As \Cref{fig:vrot_spin} shows, our spinning halo with $\lambda=\pm0.16$ follows an almost flat rotation curve, indicative of shear and epicyclic motions. By analyzing the orbits of all the halo particles, we find that the faction of the particles whose orbits are relatively circular with $\max(v_\phi)/\max(v_R)\geq 2$ and remain in the inner regions with $R\leq 5\kpc$ close to the midplane with $|z|\leq 5\kpc$ is about 0.15\% of the halo mass, corresponding to $\sim10\%$ of the disk mass at $R\leq 5\kpc$. This suggests that swing amplification in the spinning halo can still allow perturbations to grow to produce $m=2$ spirals and a weak bar, albeit much weaker than a strong bar that forms in the disk.

To summarize, our rotating halo in itself is vulnerable to the formation of non-axisymmetric structures. Small-amplitude (internal) perturbations in a spinning halo grow due to swing amplification into spirals and then to a bar which rotates in the same sense as the halo. As we will show in \autoref{sec:results}, this tendency of a spinning halo forming non-axisymmetric structures even without external perturbations promotes or delays the bar formation in a stellar disk depending on its direction of rotation relative to the halo.

\begin{figure}[t]
\centering
\plotone{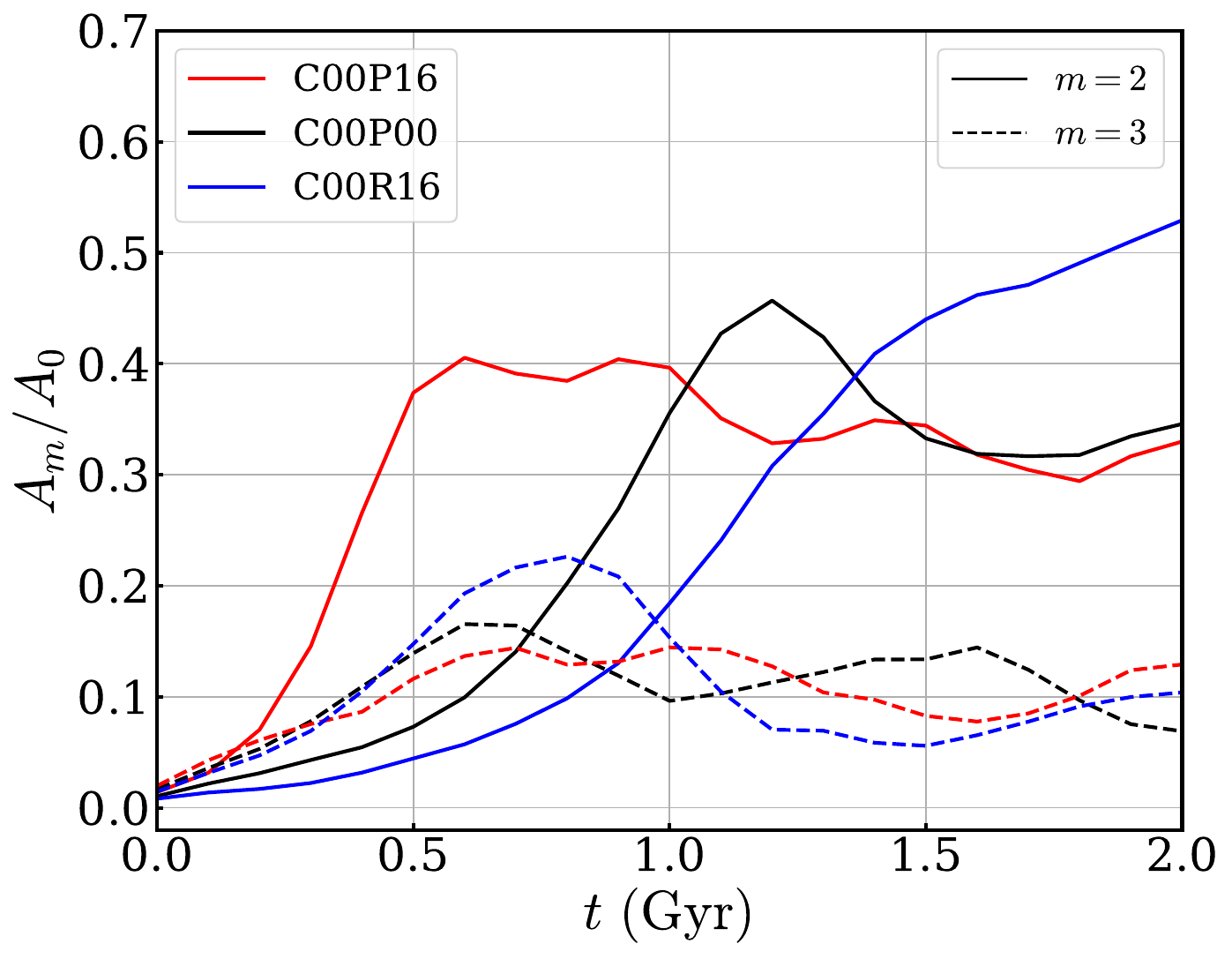}
\caption{Temporal evolution of $A_2/A_0$ (solid) and $A_3/A_0$ (dashed) for models \texttt{C00P00} (black), \texttt{C00P16} (red) and \texttt{C00R16} (blue) at early time ($t\leq2\Gyr$).
\label{fig:A2A3C00s}}
\end{figure}

\section{Full Galaxy Models} \label{sec:results}

We now present the simulation results of full galaxy models with halo and disk listed in \autoref{tbl:model}. 
We first focus on the bar formation and evolution in our models, and then explore the bar pattern speed, angular momentum transport, and vertical buckling instability.

\subsection{Bar Formation and Evolution} \label{subsec:bar}   

All of our models produce bars in the disks. The bar formation is due to successive swing amplification aided by feedback loops \citep[e.g.,][]{bnt08}: small-amplitude perturbations in a stellar disk grow as they transition from a leading to a trailing configuration, and trailing waves appear as leading waves after they pass through the galaxy center. In the growing stage, they interact gravitationally with the particles in the halo and bulge, and the halo spin certainly affects the wave growth. When the waves grow sufficiently, they shape into a bar.
\autoref{fig:snap} plots snapshots of the disk surface density for models at every $0.1\Gyr$ interval from $t=0.1$ to $1.8\Gyr$
in models \texttt{C00P00} (top), \texttt{C00P16} (middle), and \texttt{C00R16} (bottom).

We follow the method of \citet{kd18} to measure the strength of a bar that forms in the disk (see also \citetalias{jnk23}). We consider an annulus centered at radius $R$ with width $\Delta R=1\kpc$ in the regions with $R\leq 10\kpc$  of the disk, and calculate the amplitudes of the Fourier modes as
\begin{subequations}\label{e:a2b2}
\begin{align}
    a_m(R) &=  \sum_{j} \mu_j \cos{(m\phi_j)}, \\
    b_m(R) &=  \sum_{j} \mu_j \sin{(m\phi_j)},
\end{align}
\end{subequations}
where $\phi_j$ and $\mu_j$ are the azimuthal angle and mass of the $j$-th disk particle in the annulus, respectively. 
Then we define the strength of the density perturbations with mode $m$ as the maximum wave amplitude across all annuli:

\begin{equation}\label{e:barstr}
   \frac{A_m}{A_0} = \max_{R}\left\{ \frac{[a_m(R)^2+b_m(R)^2]^{1/2}}{\sum_{j}\mu_j}\right\}.
\end{equation}
Note that the term inside the curly brackets in \cref{e:barstr} is equal to $|A(m,0)|$ defined in \cref{e:logarm} if the particles have the same mass.

\autoref{fig:A2A3C00s} plots the temporal evolution  of $A_2/A_0$ and $A_3/A_0$ at early time ($t\leq 2\Gyr$) for models \texttt{C00P00} (black), \texttt{C00P16} (red), and \texttt{C00R16} (blue). 
In model \texttt{C00P00} with no halo spin, the disk favors the growth of $m=2$ and 3 spirals, with the $m=3$ mode dominating at $t\lesssim 0.8\Gyr$ (see the top row of \autoref{fig:snap}). These modes with different $m$ interact nonlinearly with each other, and one arm of the $m$ = 3 spirals merges with the other two arms, producing a bar at $t\sim0.9\Gyr$ \citep[see also,][]{seo19}. The $m=3$ mode in these models decays secularly to become $A_3/A_0 \sim 3 \times 10^{-2}$ at the end of the runs ($t=10\Gyr$). 

In model \texttt{C00P16} with $\lambda=0.16$, the growth of the $m=2$ mode in the disk is assisted by the prograde halo which has an intrinsic tendency of amplifying the $m=2$ perturbations rotating in the same sense as the halo. Consequently, the $m=2$ spirals in this model become stronger and thus transform to a bar earlier ($t\sim0.5\Gyr$) than in model \texttt{C00P00}  (see the middle row of \autoref{fig:snap}). We note that the regions outside the bar in model \texttt{C00P16} are more strongly perturbed by the $m=2$ spirals than in model \texttt{C00P00}, limiting further growth of the bar. In model \texttt{C00R16} with $\lambda=-0.16$, in contrast, the growth of the $m=2$ mode in the disk is opposed to some extent by the counter-rotating halo, which in turn allows for the $m=3$ mode to grow more strongly than in model \texttt{C00P00}. Therefore, it takes longer for the $m=3$ spirals to transform eventually into a bar than under the non-spinning halo. 

\begin{figure} 
\centering
\plotone{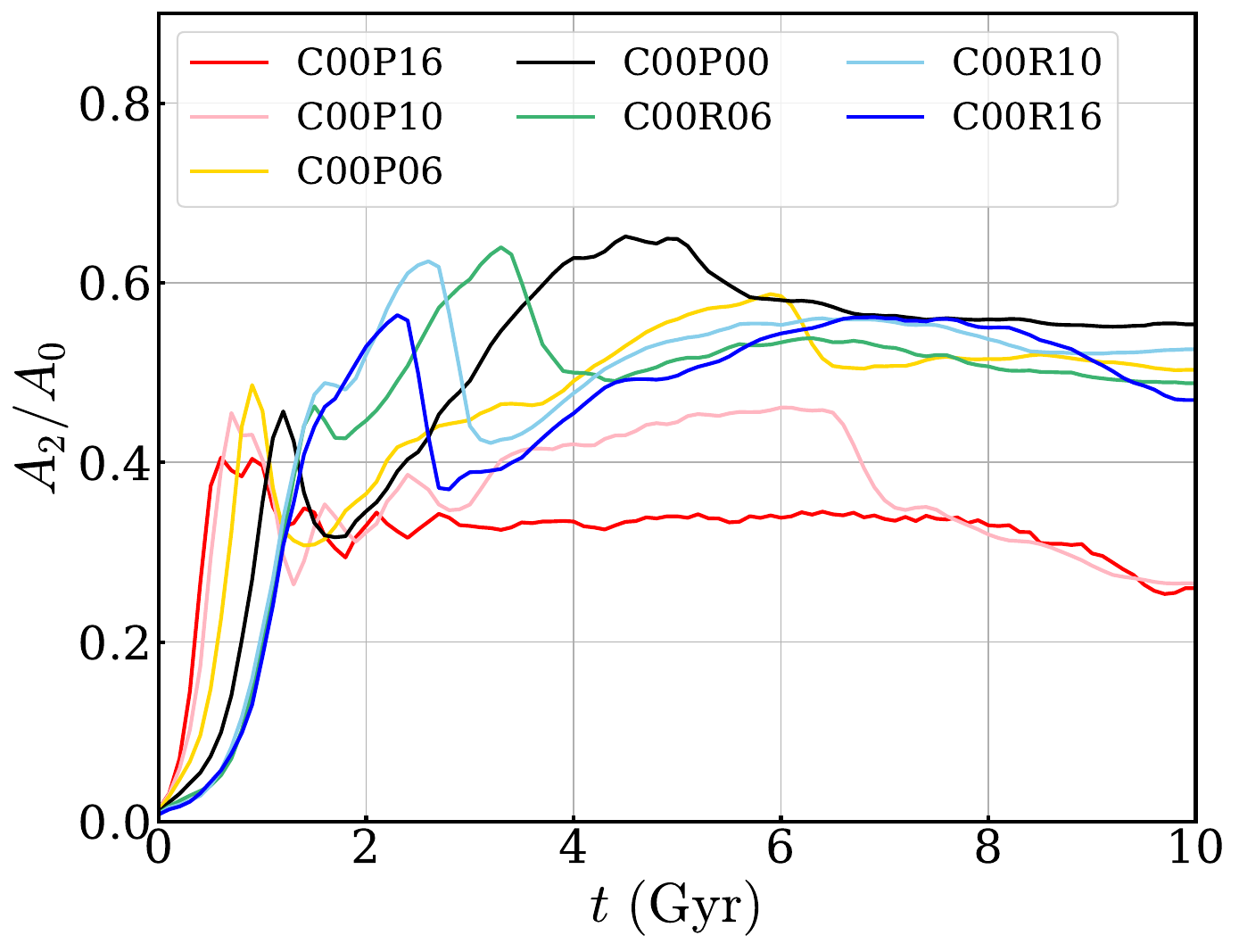}
\plotone{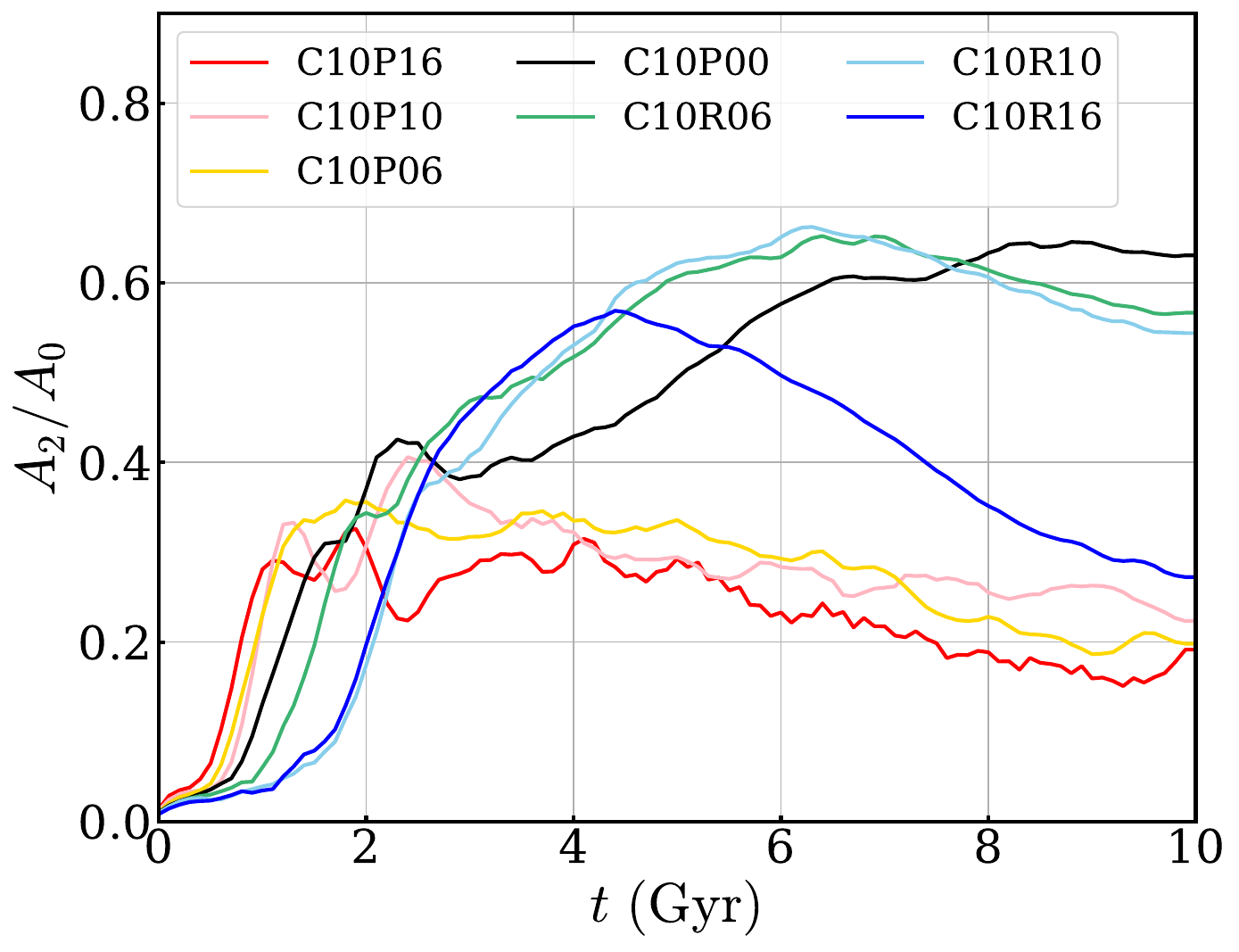}
\caption{Temporal evolution of the bar strength $A_2/A_0$ 
for the \texttt{C00} series (top) and \texttt{C10} series (bottom) for $t\leq 10\Gyr$. A bar tends to form early under a prograde halo and without bulge. 
\label{fig:A2}}
\end{figure}

\begin{figure*}
\centering
\epsscale{1.2}
\plotone{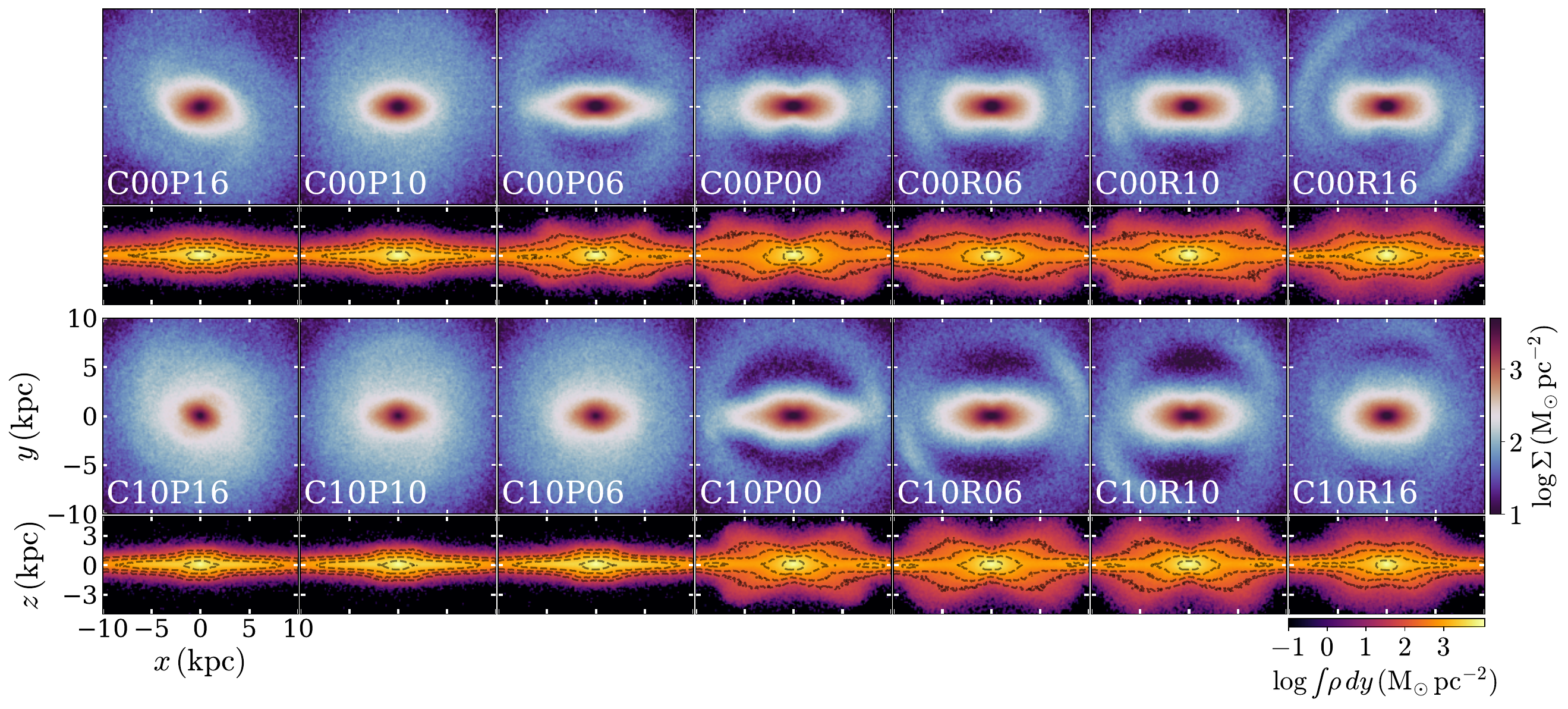} 
\caption{Snapshots of the disk surface density (top) and the projected disk density along the bar semi-minor axis (bottom) at the end of the runs ($t = 10 \Gyr$) for all models. The upper and lower rows correspond to the models in the \texttt{C00} and \texttt{C10} series, respectively. The $x$- and $y$-axis correspond to the semi-major and semi-minor axis, respectively. In the edge-on views, the dotted contours denote $\int \rho \,dy=10^{3.5}, 10^{3.0}, 10^{2.5}, 10^{2.0}\Msun\pc^{-2}$ from inside to outside.
\label{fig:snapAll}}
\end{figure*}

\autoref{fig:A2}  plots the temporal evolution of $A_2/A_0$ for \texttt{C00} series (upper panel) and \texttt{C10} series (lower panel) over entire evolution.  The bar in model \texttt{C00P16} stops growing at $t\sim1.0\Gyr$. At this time, the inner parts of the spirals dominating the outer regions become in phase with the bar, making $A_2/A_0$ increase temporarily to $\sim0.4$. As the spirals rotate relative to the bar and become out of phase, $A_2/A_0$ decreases to  $\sim 0.32$ at $t\sim2\Gyr$. Since the regions outside the bar are strongly perturbed by the $m=2$ spirals, they do not participate in the bar growth, making $A_2$ stay almost constant for a long period. The bar undergoes a weak buckling instability at $t\sim9.5\Gyr$ (see \autoref{subsec:buckling}).

In model \texttt{C00P00} with no halo spin, $A_2/A_0$ peaks near $t\sim{1.1}\Gyr$ when the spirals are in phase with the bar, after which $A_2/A_0$ decreases as the bar becomes out of phase with the spirals. Since the regions outside the bar possess weaker spirals and are thus less disturbed than in model \texttt{C00P16}, the bar in this model can grow in size and strength by gathering outer particles until it undergoes a buckling instability at $t\sim 5.2\Gyr$.  In model \texttt{C00R16} with a counter-rotating halo, the bar grows slower but more strongly than the other models since the spiral perturbations in the outer regions are weakest. The bar becomes so strong that the buckling instability occurs at $t\sim 2.5\Gyr$, earlier than the other models.

The models in the \texttt{C10} series with a classical bulge also form a bar, but later than the no-bulge counterparts. This is because the gravity of a bulge reduces the feedback loop and swing amplification. The bars in the \texttt{C10} series are weaker and shorter at early time ($t\lesssim2\Gyr$) than those in the \texttt{C00} series, indicating that the presence of a non-rotating bulge, in general, resists the bar formation (\citealt{sell80,bnt08,kd18,se18}; \citetalias{jnk23}). 

The bars in models \texttt{C10P16}, \texttt{C10P10} and \texttt{C10P06} are always weaker than those in the no-bulge counterparts. However, the bars in the other models with a bulge do not undergo buckling instability and can thus be stronger than those without a bulge. For example, the bar in model \texttt{C10P00} keeps growing to reach $A_2/A_0\sim 0.64$, while the bar in model \texttt{C00P00} suddenly becomes weaker after experiencing buckling instability at $t\sim 5.2\Gyr$.

\autoref{fig:snapAll} plots the snapshots of the disks projected along the vertical direction and the bar semi-minor axis at $t=10\Gyr$ for all models. The bar rotating in the counterclockwise direction is oriented so that the $x$- and $y$-axes correspond to the semi-major and semi-minor axes, respectively. The bar in model \texttt{C00P16} shows a twist of isodensity contours in the outer parts. The bar in model \texttt{C10P16} is so weak that it can be regarded as an oval. The surface density near the center has a dumbbell-like distribution only in model \texttt{C00P00}. The models with relatively strong bars (other than models \texttt{C00P16}, \texttt{C00P10}, \texttt{C10P16}, \texttt{C10P10} and \texttt{C10P06}) possess a \ac{BPS} bulge as well as an inner ring just outside the bar. 

\begin{figure}
\centering
\plotone{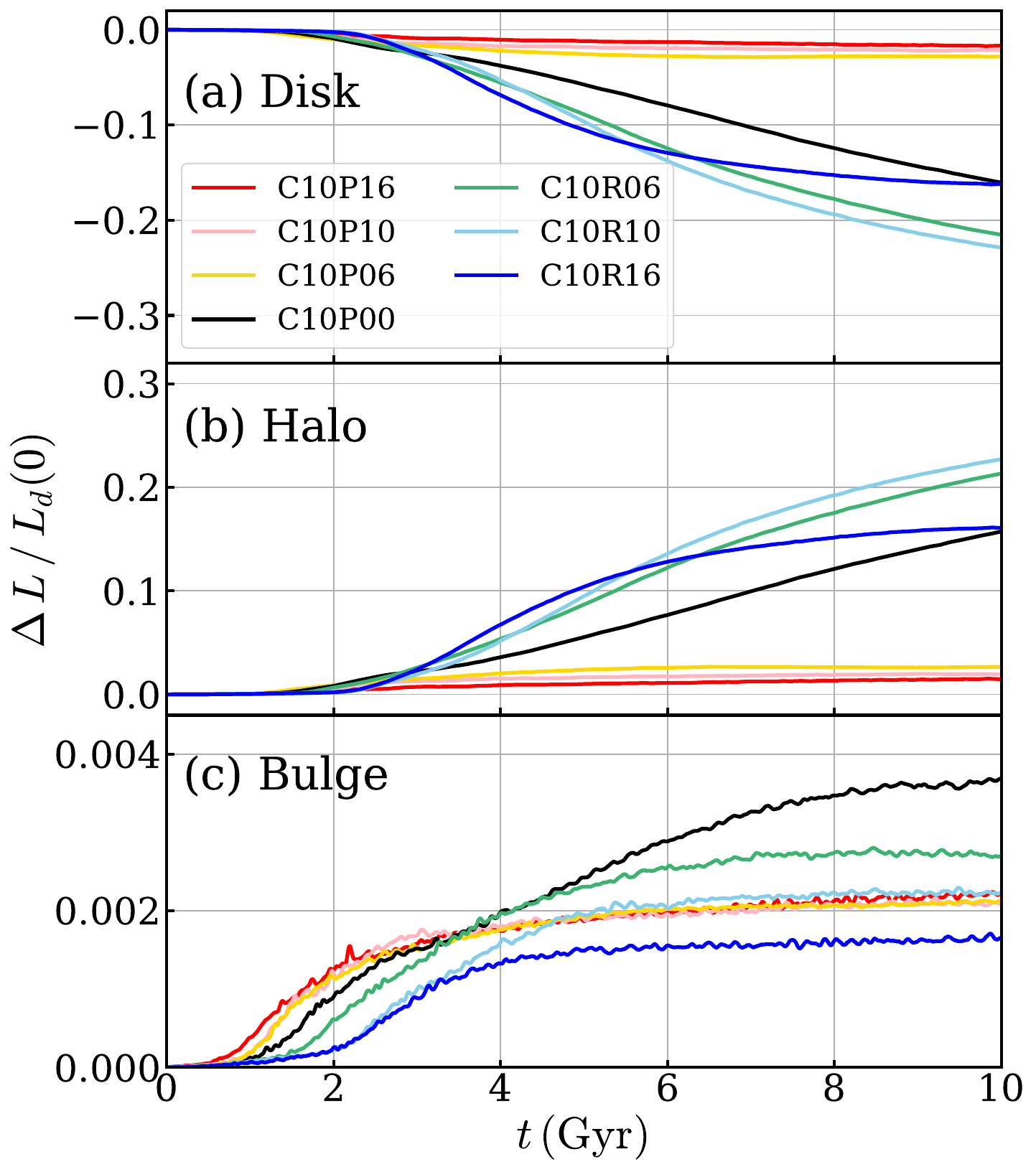}
\caption{Temporal changes in the angular momentum $\Delta L\equiv L(t)-L(0)$ of the entire disk (top), halo (middle), and bulge (bottom) for the models in the \texttt{C10} series.  The ordinate is normalized by the initial angular momentum of the entire disk.
\label{fig:Lz}}
\end{figure}

\subsection{Angular Momentum and Bar Pattern Speed} \label{subsec:omega}  

After formation, a bar embedded in a disk transfers its angular momentum to a halo (and a bulge). To study the radial and temporal dependence of the angular momentum transfer, we bin each component radially into concentric cylindrical shells with width $\Delta R=1\kpc$ and calculate the vertical component of the angular momentum 
\begin{equation}
   L = \sum_{i}\mu_i (x_i v_{y,i} - y_i v_{x,i})
\end{equation}
in each shell.  \autoref{fig:Lz} plots $\Delta L\equiv L(t)-L(0)$ summed over the entire shells as a function of time for the models in the \texttt{C10} series.  Overall, the amount of angular momentum absorbed by the bulge is much less than that by the halo. The disk under a counter-rotating halo loses more angular momentum than under the prograde halo, making the bar in the former grow stronger and longer at the end of the runs (see \autoref{fig:snapAll}).  Of course, the actual temporal change of $\Delta L$ depends on not only $\lambda$ but also the bar properties such as strength, length, pattern speed, mass distribution, etc.

\autoref{fig:Lzmap} displays the rate of angular momentum transfer, $dL/dt$, as a function of $R$ and $t$ in the halo and disk for models \texttt{C00P16}, \texttt{C00P00}, and \texttt{C00R16}. 
When $\lambda>0$, the halo particles inside the corotation resonance $R_\textrm{CR}$ with the bar rotate faster than, and thus lose their angular momenta to the bar, while the particles in the outer halo gain angular momentum \citep{long14,col18}. Since $R_\textrm{CR}$ is larger for larger $\lambda$, the angular momentum transfer from disk to halo tends to be larger for smaller $\lambda>0$. In model \texttt{C00P16}, most angular momentum transfer occurs near $t\sim1\Gyr$ when the bar is strongest. The bar angular momentum is also transferred to the outer disk at $R>R_\textrm{CR}$.

When $\lambda\leq0$, in contrast, all the halo particles lag behind the bar and thus allow the efficient angular momentum transfer. \autoref{fig:Lzmap} shows that the halo in model \texttt{C00P00} absorbs angular momentum mainly through three resonances which extend radially outward as the bar grows in size over time (e.g., \citealt{vil09,col18,col19b,col19a}). Again, the largest $dL/dt$ occurs near $t\sim3$--$5\Gyr$ when the bar is strongest. The outer disk absorbs the angular momentum emitted by the bar. The halo in model \texttt{C00R16} absorbs angular momentum strongly at $t\sim1$--$3\Gyr$ and $\sim5$--$7\Gyr$, corresponding to a strong bar before the buckling instability and in the regrowth stage (see \autoref{fig:A2}). 

\begin{figure*}
\centering
\plotone{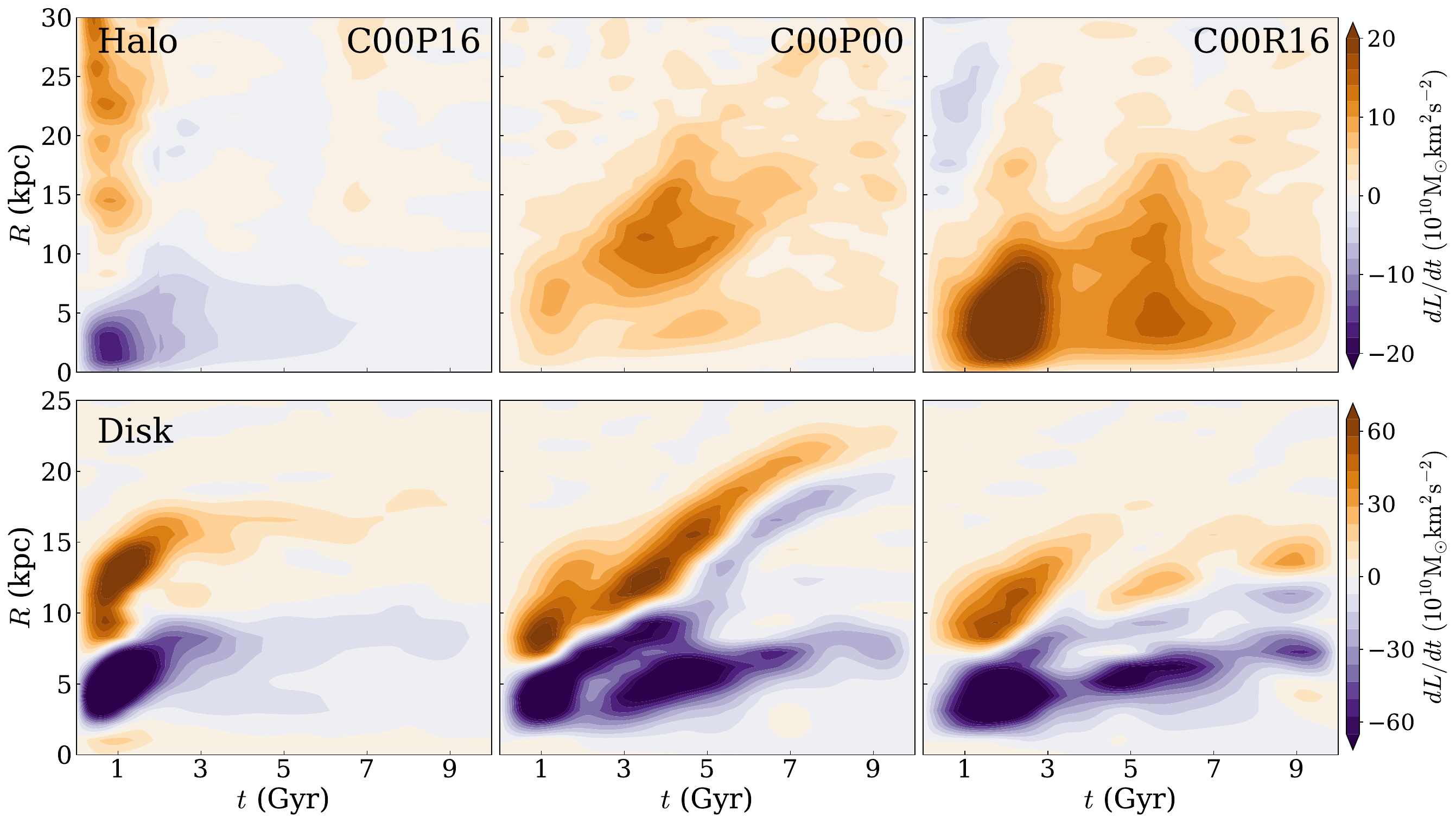}
\caption{ Rates of the angular momentum change, $dL/dt$, as a function of cylindrical radius and time for models \texttt{C00P16} (left), \texttt{C00P00} (middle), and \texttt{C00R16} (right). The cylindrical shells have a width $\Delta R =1 \kpc$ and height $|z|=\infty$ for the halos (upper panels) and the stellar disks (lower panels).  
\label{fig:Lzmap}}
\end{figure*}

We calculate the bar pattern speed, $\Omega_b$, using the cross-correlation of the disk surface density in the annular regions with width $\Delta R = 0.1 \kpc$ at $R=2\kpc$: we check that the resulting $\Omega_b$ agrees within $\sim1\%$ with that from the method that utilizes the temporal changes in the bar position angle $\psi=0.5\tan^{-1}(b_2/a_2)$ \citepalias{jnk23}. 
\autoref{fig:omega} plots the temporal changes in the bar pattern speed for all models. As the bar loses its angular momentum to the halo and bulge, its pattern speed decreases over time. The initial bar pattern speed tends to be higher in models with a prograde halo and with a bulge. This is consistent with the numerical results that a short bar tends to rotate fast (\citealt{agu15,roshan21,lee22,frankel22}; \citetalias{jnk23}).
Note that the bar slows down very slowly in models with $\lambda=0.16$ since the angular momentum absorption by the halo outside $R_\mathrm{CR}$ is almost balanced by the angular momentum emission by the halo inside $R_\mathrm{CR}$. 
Models with smaller $\lambda \;(>0)$ have smaller $R_\mathrm{CR}$ and thus causes $\Omega_b$ to decay faster over time. This is qualitatively consistent with the analytic result of \citet{chi24} who showed that the dynamical friction of a bar due to halo particles weakens with increasing halo spin.  Models with $\lambda<0$ have $\Omega_b$ decreasing more or less similarly to that in the $\lambda=0$ models, largely consistent with the result of \citet{col19b} in that the amount of angular momentum transport is insensitive to $\lambda$ as long as the halo has a retrograde spin.

\subsection{Buckling Instability}\label{subsec:buckling}

All the bars in our models thicken vertically over time, evolving into a \ac{BPS} bulge if strong. The \ac{BPS} strength is usually defined as 
\begin{equation}\label{e:Ps}
P_s = \max_R\left(\frac{\widetilde{|z|}}{\widetilde{|z_0|}}\right),
\end{equation}
where the tilde denotes the median and $z_0$ is the initial height for the disk particles at $R\leq10\kpc$ \citep{ian15,frag17,seo19}. \autoref{fig:Ps} plots temporal evolution of $P_s$ for all models. Note that the models in the \texttt{C00} series with no bulge (except for model \texttt{C00P16}) go through a short period where $P_s$ increases rapidly due to buckling instability. In contrast, $P_s$ in the models in the \texttt{C10} series with a bulge does not experience such a rapid increase. Instead, it increases relatively slowly due to the vertical heating of disk particles via gravitational interactions with the bar without undergoing buckling instability.

\autoref{fig:xz} compares the projected disk densities along the bar semi-minor axis in models \texttt{C00P00} (left) and \texttt{C10P00} (right) for $t=4.5$--$7.0\Gyr$. Clearly, the bars in both models produce a \ac{BPS} bulge. Note that the projected disk densities in model \texttt{C00P00} become asymmetric with respect to the $z=0$ plane at $t\sim5.5\Gyr$, indicative of the buckling instability. However, the bar in model \texttt{C10P00} almost always maintains a mirror symmetry about the midplane, while it gradually thickens vertically.  This relatively slow thickening of the bars can be caused by 2:1 and 4:1 vertical resonances \citep{comb90,pfe91,pat02,sg20}. This indicates that the buckling instability is not a necessary but sufficient condition for forming a \ac{BPS} bulge \citep{li23}.

\begin{figure}[t]
\centering
\plotone{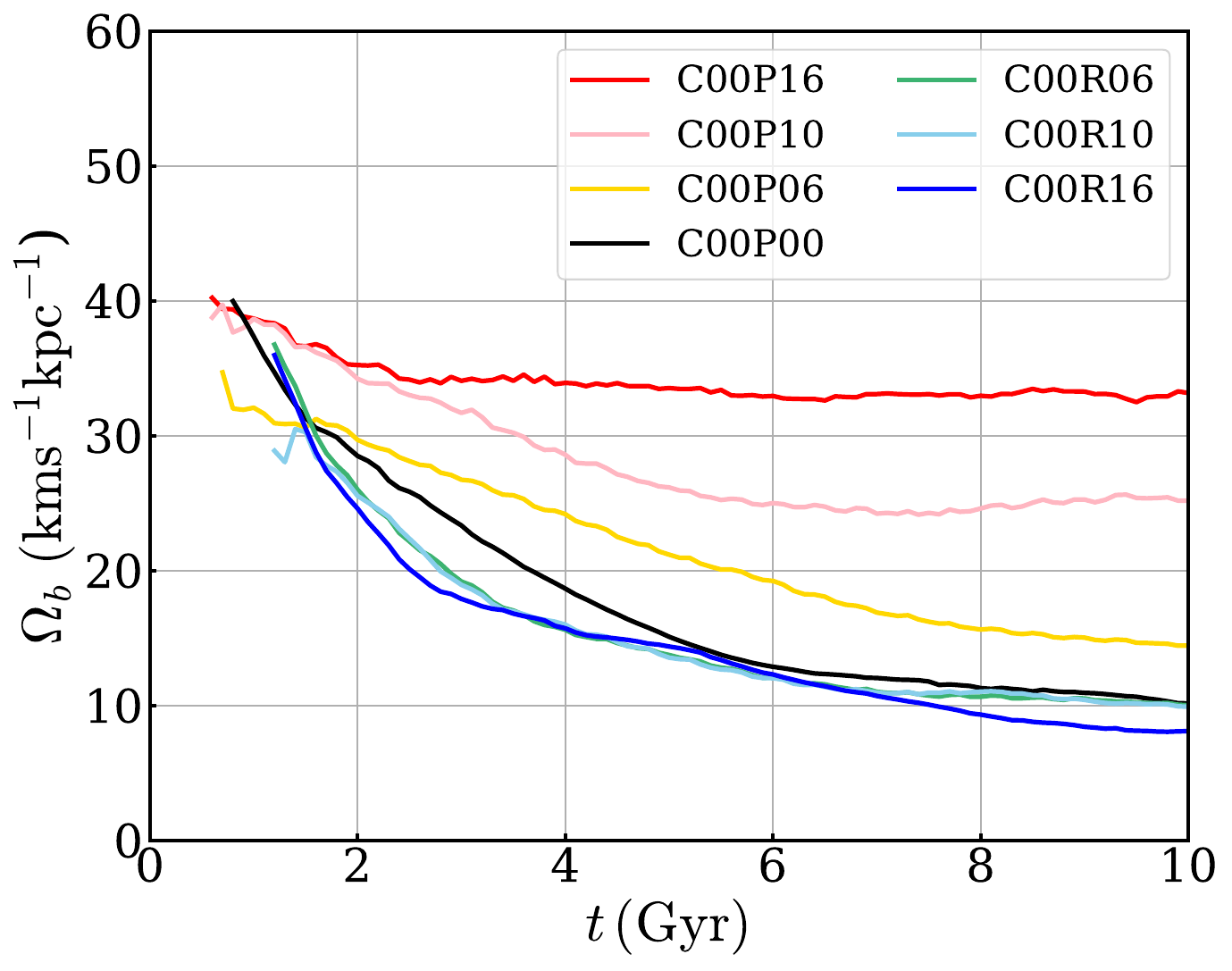}
\plotone{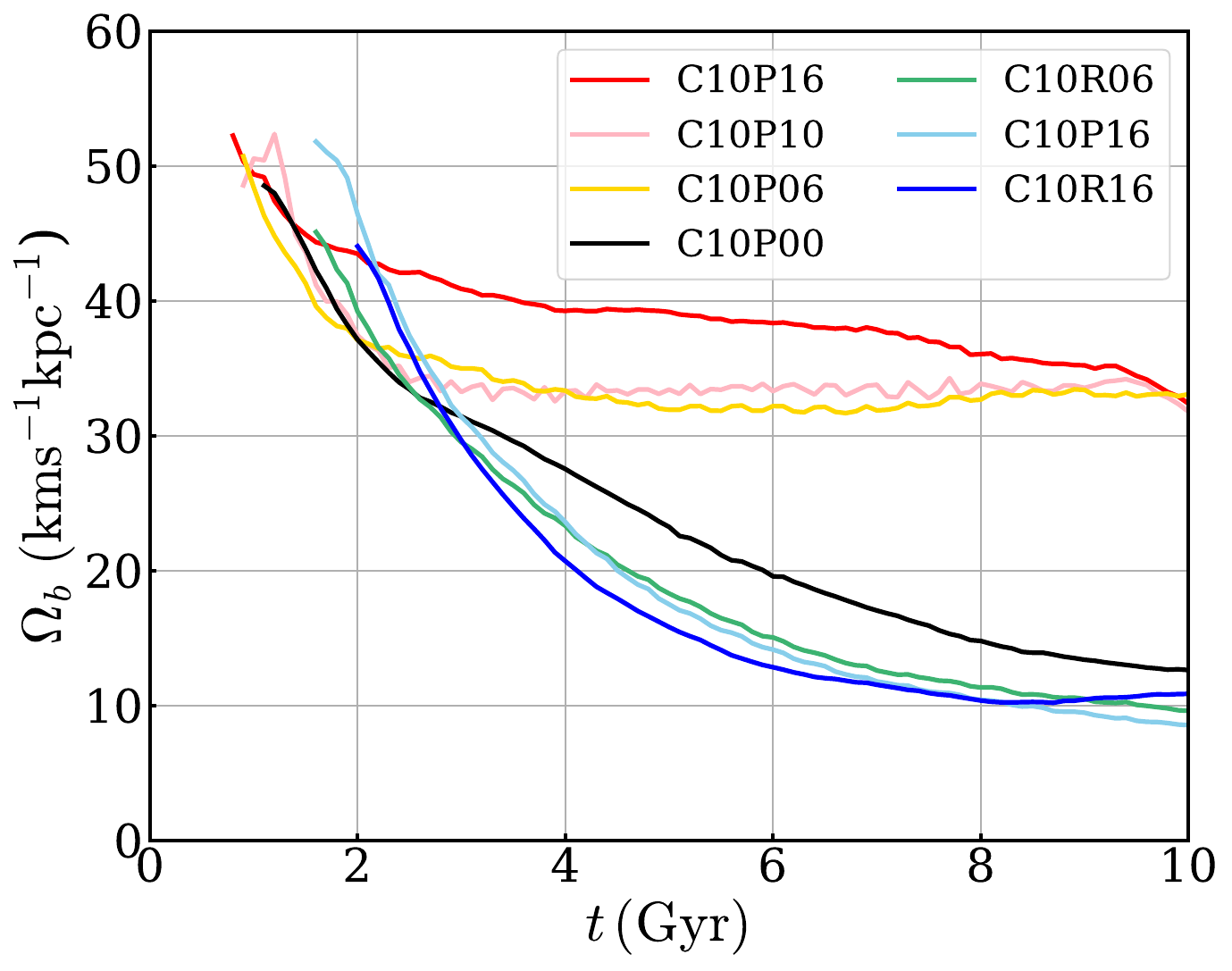}
\caption{Temporal changes of the bar pattern speed $\Omega_{b}$ 
for the \texttt{C00} series (top) and \texttt{C10} series (bottom). \label{fig:omega}}
\end{figure}

\begin{figure}[ht]
    \centering
    \plotone{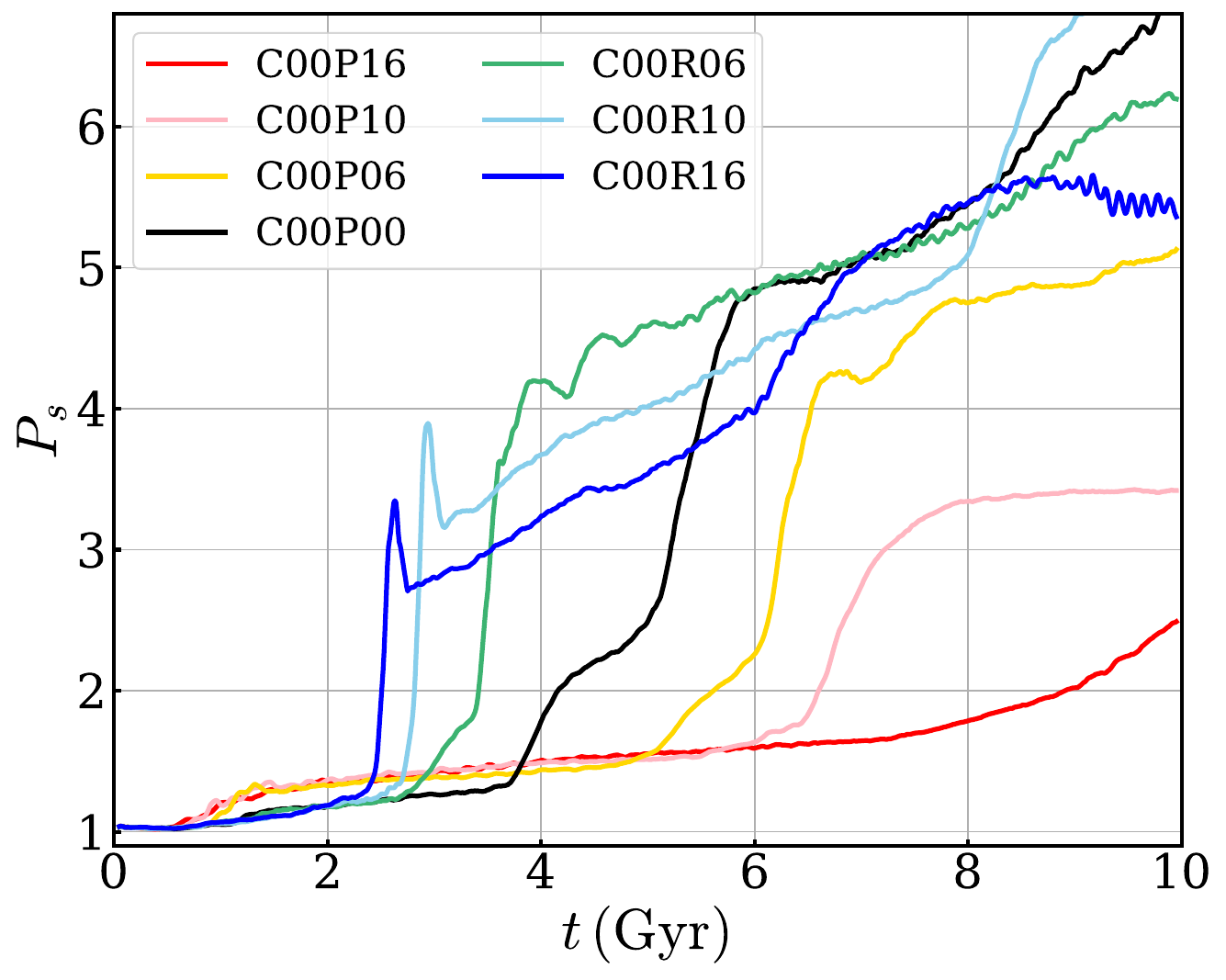}
    \plotone{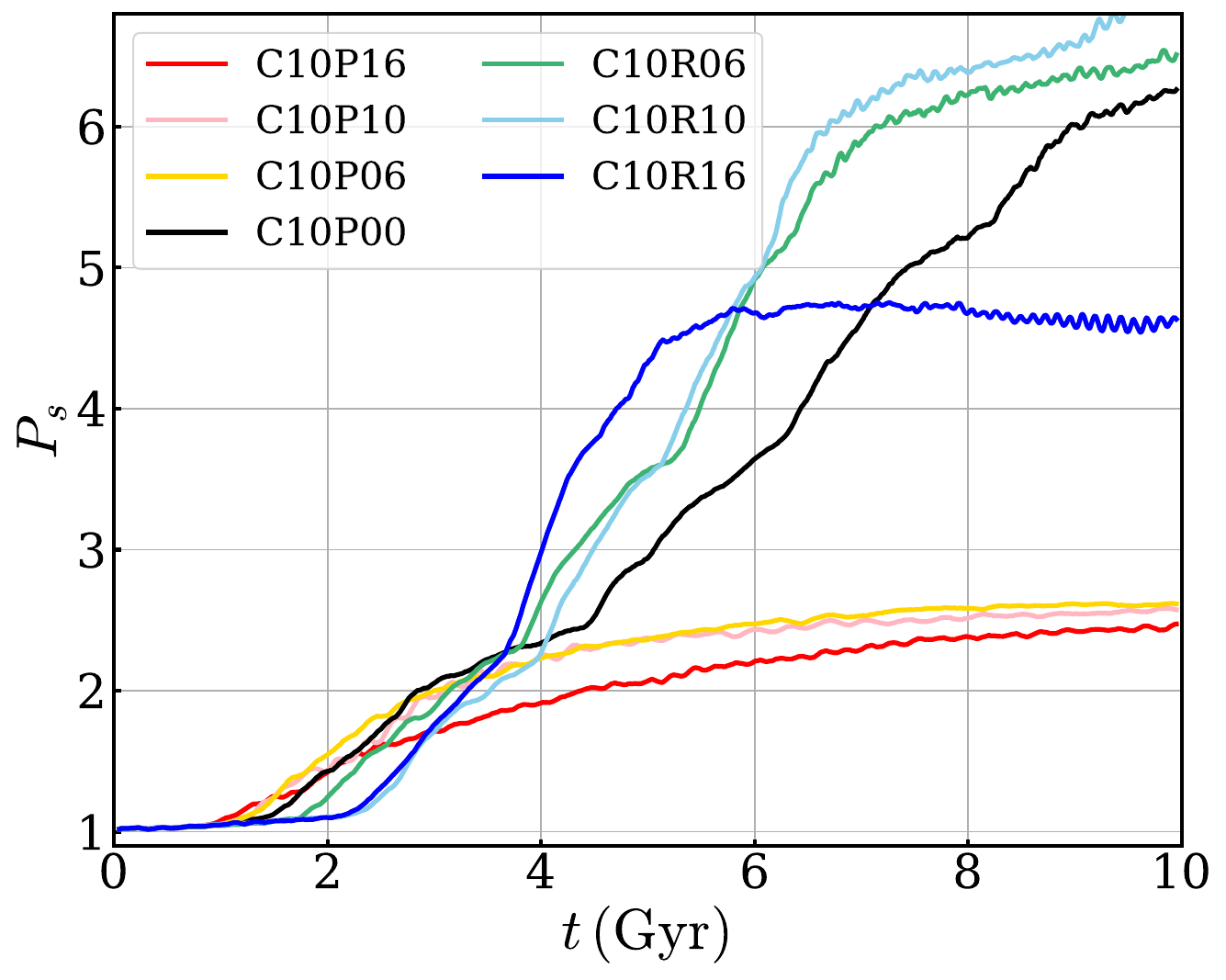}
    \caption{Time evolution of the \ac{BPS} strength, $P_s$, defined in \cref{e:Ps} for the models in the \texttt{C00} series with no bulge (top) and \texttt{C10} series with a bulge (bottom).}
    \label{fig:Ps}
\end{figure}

\begin{figure}[t]
\centering
\plotone{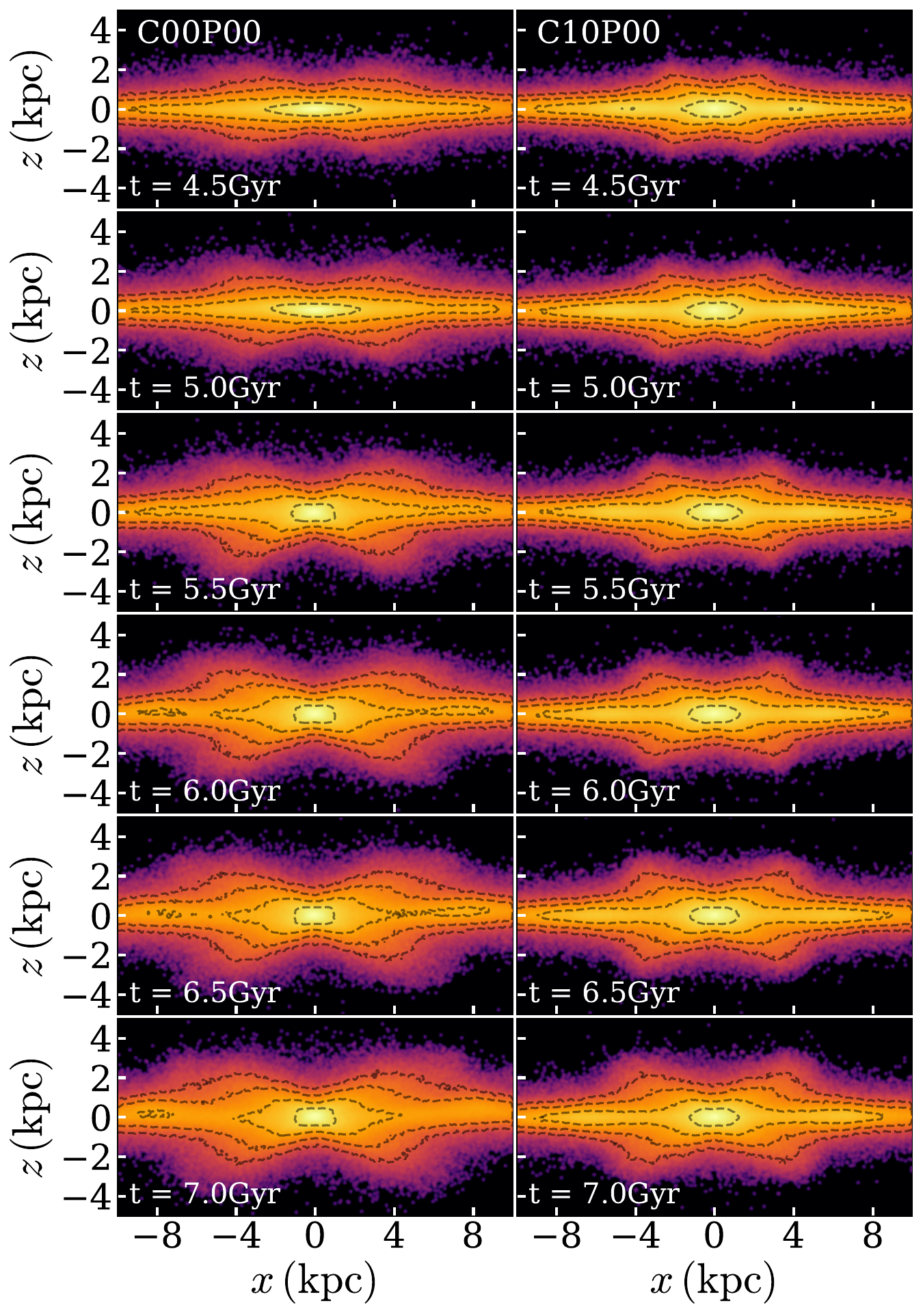} 
\caption{Contours of the logarithm of the projected disk densities at $t = 4.5$--$7.0\Gyr$ for models \texttt{C00P00} (left) and \texttt{C10P00} (right). 
The buckling instability occurs at $t \sim 5.5 \Gyr$ in model \texttt{C00P00}. The $x$- and $z$-axis correspond to the bar semi-major and vertical axes, respectively. The dotted contours denote $\int \rho \,dy=10^{3.5}, 10^{3.0}, 10^{2.5}, 10^{2.0}\Msun\pc^{-2}$  from inside to outside.
\label{fig:xz}}
\end{figure}

\begin{figure}[ht]
\centering
\plotone{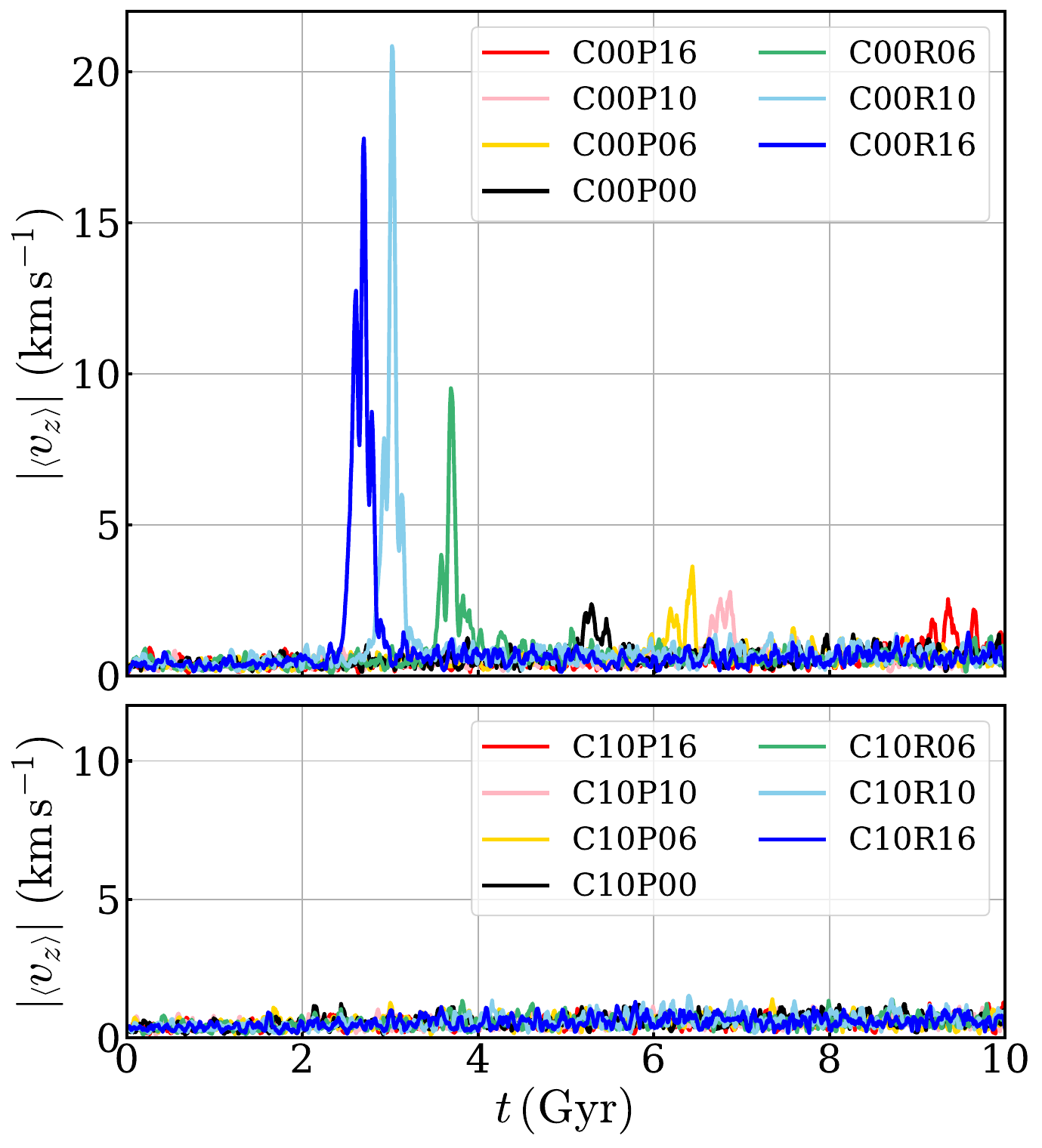}
\caption{Temporal evolution of the mean vertical velocity $|\langle v_z\rangle|$ at $R=2.0 \kpc$ of the disk for the models in the \texttt{C00} series (top) and the \texttt{C10} series (bottom). 
\label{fig:vz}}
\end{figure}

\begin{figure*}[ht]
\centering
\plotone{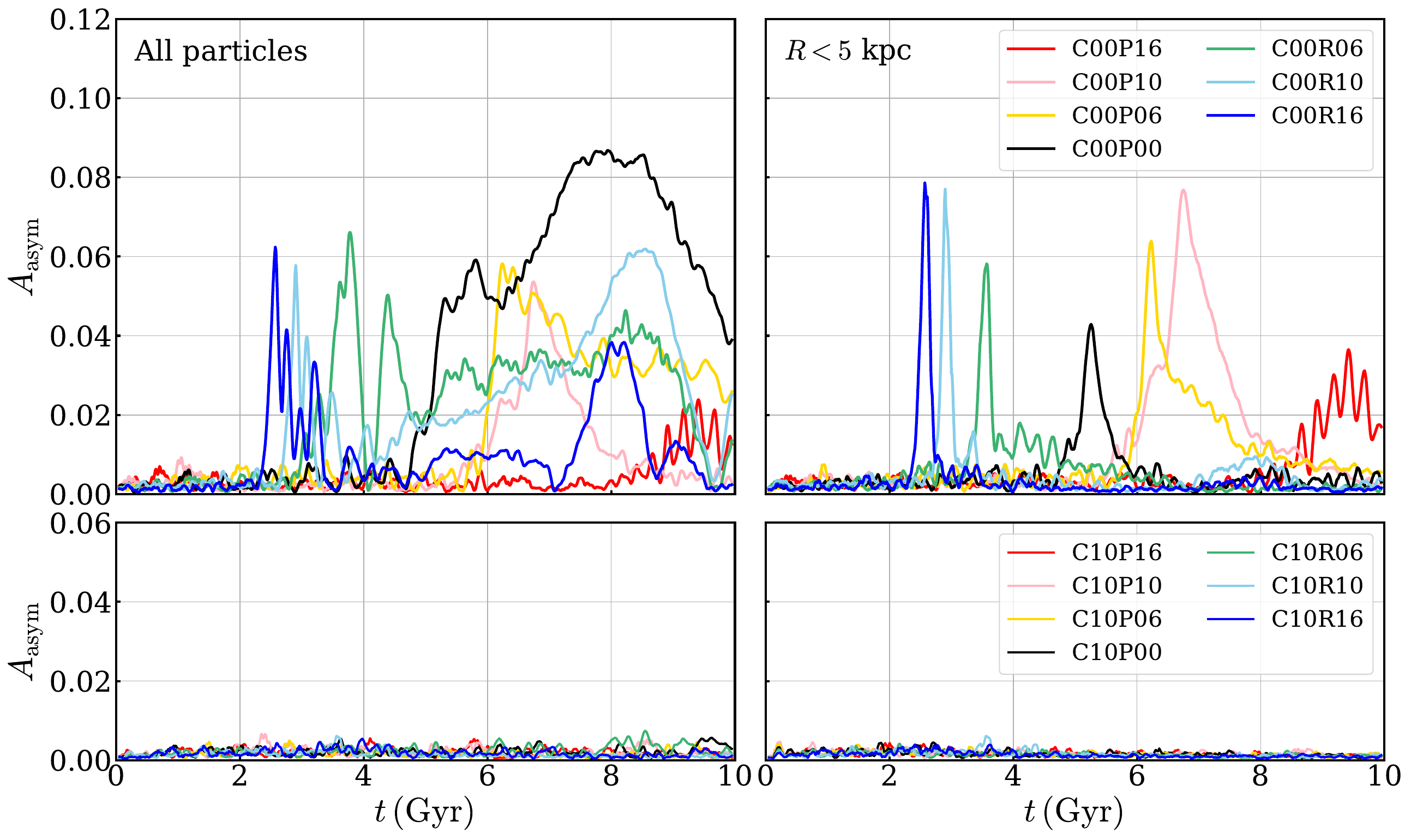}
\caption{Temporal evolution of the vertical asymmetry parameter $A_\mathrm{asym}$, defined in \cref{e:Asym}, from all the disk particles  (left) and the particles at $R < 5 \kpc$ (right) for the models in the \texttt{C00} series (top) and the \texttt{C10} series (bottom).\label{fig:Asym}}
\end{figure*}

Since the buckling instability deforms the disk in the vertical direction, it naturally involves the asymmetry in the vertical velocities of disk particles. One way to quantify the onset and strength of the buckling instability is to monitor the evolution of $|\langle v_z\rangle|$, where the angle brackets denote the spatial average at $R=2\kpc$ \citep{kwak17,kwak19}. \autoref{fig:vz} plots $|\langle v_z\rangle|$ against time for all models. A sharp peak in $|\langle v_z\rangle|$ corresponds to the buckling instability, which starts to occur at $t=2.5, 2.9, 3.7, 5.2, 6.2, 6.8 \Gyr$ for models \texttt{C00R16}, \texttt{C00R10}, \texttt{C00R06}, \texttt{C00P00}, \texttt{C00P06}, \texttt{C00P10}, respectively. 
Model \texttt{C00P16} also undergoes a buckling instability at $t\sim9.5\Gyr$, although it is very mild. The bottom panel of \autoref{fig:vz} shows that $|\langle v_z\rangle|$  of the models in the \texttt{C10} series exhibit only noises, suggesting that they do not experience buckling instability at all.

Another way to measure the strength of the buckling instability is to use the vertical asymmetry parameter of the disk introduced by \citet{smir18} as 
\begin{equation}\label{e:Asym}
A_\textrm{asym} = \left|\frac{A_2(z>0)-A_2(z<0)}{A_0}\right|,
\end{equation}
where $A_2(z>0)$ and $A_2(z<0)$ denote the Fourier amplitudes of the $m=2$ mode applied to the disk particles above and below the $z=0$ plane, respectively \citep[see also][]{li24}. \autoref{fig:Asym} plots $A_\textrm{asym}$ from all the disk particles (left) as well as the particles located at $R\leq 5\kpc$ (right). The models with no bulge have $A_\textrm{asym}$ varying a lot over time, while $A_\textrm{asym}$ for the models with a bulge is close to zero. For the models without a bulge, $A_\textrm{asym}$ from the entire disk exhibits longer-term variations than that from the regions with $R\leq 5\kpc$, suggesting that the buckling instability initiates from smaller radii and propagates radially outward. Compared to $|\langle v_z\rangle|$, $A_\textrm{asym}$ for the models in the \texttt{C00} series are more widely distributed over time. This is because $A_\textrm{asym}$ captures the asymmetry in a wider radial range and the buckling instability occurs in the inner disk first and then propagates outward \citep{mar06}.  Note that $|\langle v_z\rangle|$ can be small if the nodes of the vertical bending in the course of buckling are located close to $R=2\kpc$, as in models \texttt{C00P00}, \texttt{C00P06}, and \texttt{C00P10}. This suggests that the peak values of $A_\textrm{asym}$ represent the overall 
strength of the buckling instability better than $|\langle v_z\rangle|$.

As \cref{fig:vz,fig:Asym} show, the buckling instability occurs earlier in our models with smaller $\lambda$, that is, as the fraction of retrograde halo particles increases. This trend is not observed in previous investigations \citep{long14, col18,col19b,col19a,kns22}.
In \cite{col18,col19b}, the maximal pre-buckling amplitude of $A_2/A_0$ is almost insensitive to $\lambda>0$, while it becomes smaller as $\lambda$ decreases in the models with a retrograde halo. In our models, the pre-buckling strength is most significant in the models with $\lambda=0$ and decreases as $|\lambda|$ increases in models with either a prograde or retrograde halo (see \autoref{fig:A2}).

It is well known that the buckling instability occurs when the radial velocity dispersion $\sigma_R$ of disk particles exceeds a critical value relative to the vertical velocity dispersion $\sigma_z$ \citep[e.g.,][]{bnt08}.  \cite{toomre66} and \cite{ara87} found that the critical value is at $\sigma_z/\sigma_R \sim 0.3$ for non-rotating, razor-thin disks. For realistic disks with spatially-varying $\sigma_z/\sigma_R$, \cite{raha91} found that the disks undergo buckling instability when $\sigma_z/\sigma_R \lesssim 0.25$--$0.55$. For barred galaxies, $N$-body simulations of \citet{mar06} and \citet{kwak17} found that buckling occurs when $\sigma_z/\sigma_R \lesssim 0.6$. 

\autoref{fig:sigma} plots the temporal changes in $\sigma_z/\sigma_R$ of the disk particles at $R = 2 \kpc$ for the models in the \texttt{C00} series (top) and the \texttt{C10} series (bottom). For most of the models without a bulge, the onset of the buckling instability corresponds to $\sigma_z/\sigma_R\simeq 0.47$--$0.60$. Note that model \texttt{C00P16} has $\sigma_z/\sigma_R\sim0.75$ at $t=9.5\Myr$, but it still exhibits mild buckling instability, as evidenced by the asymmetry parameter shown in \autoref{fig:Asym}.  All the models with a bulge have $\sigma_z/\sigma_R\gtrsim 0.55$, and remain stable against buckling instability. Instead, the bars in these models thicken gradually to produce \ac{BPS} bulges (see \autoref{fig:xz}). This indicates that the critical value of $\sigma_z/\sigma_R$ for buckling instability is applicable only to the models without a bulge.
\begin{figure}[t]
\centering
\plotone{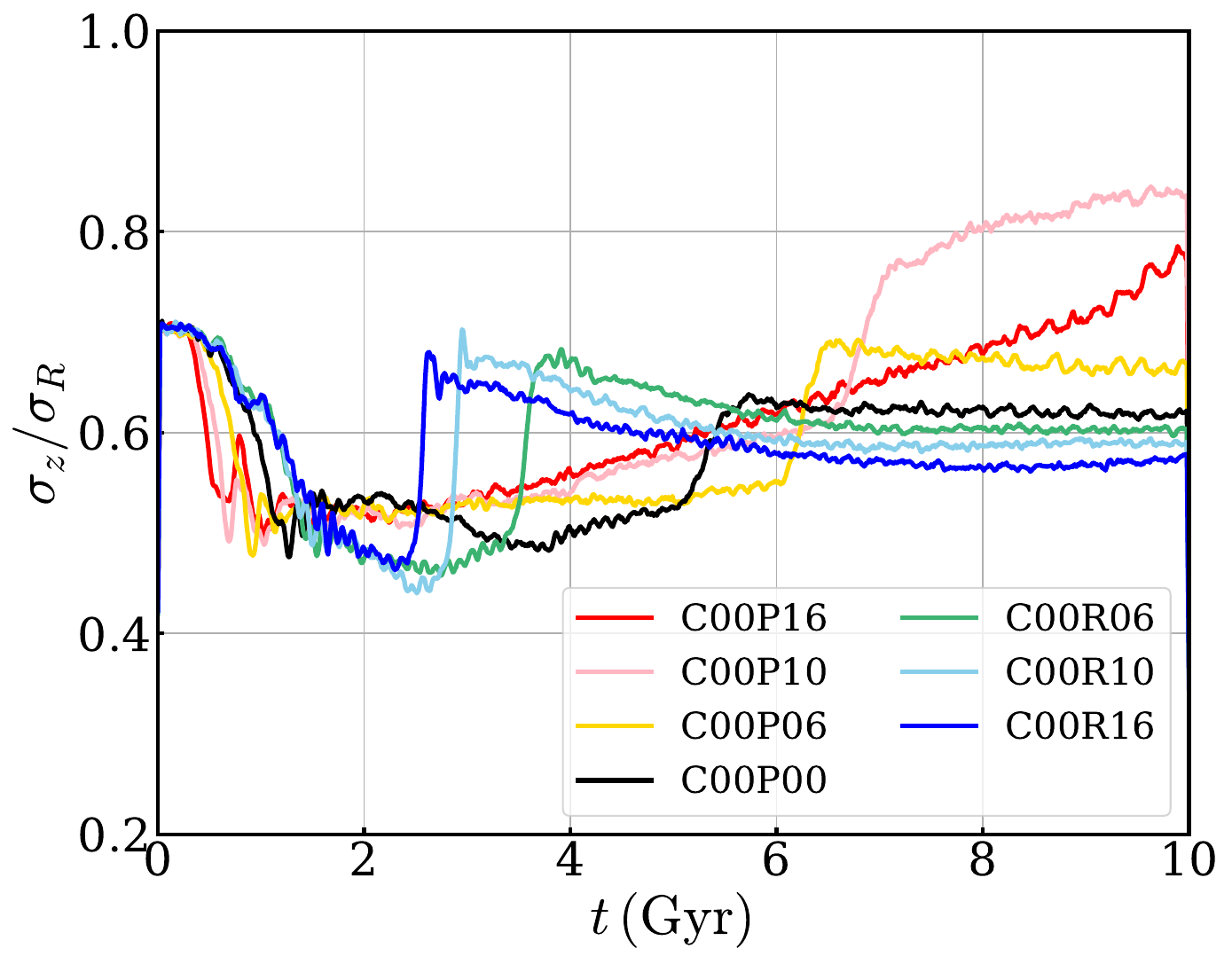} 
\plotone{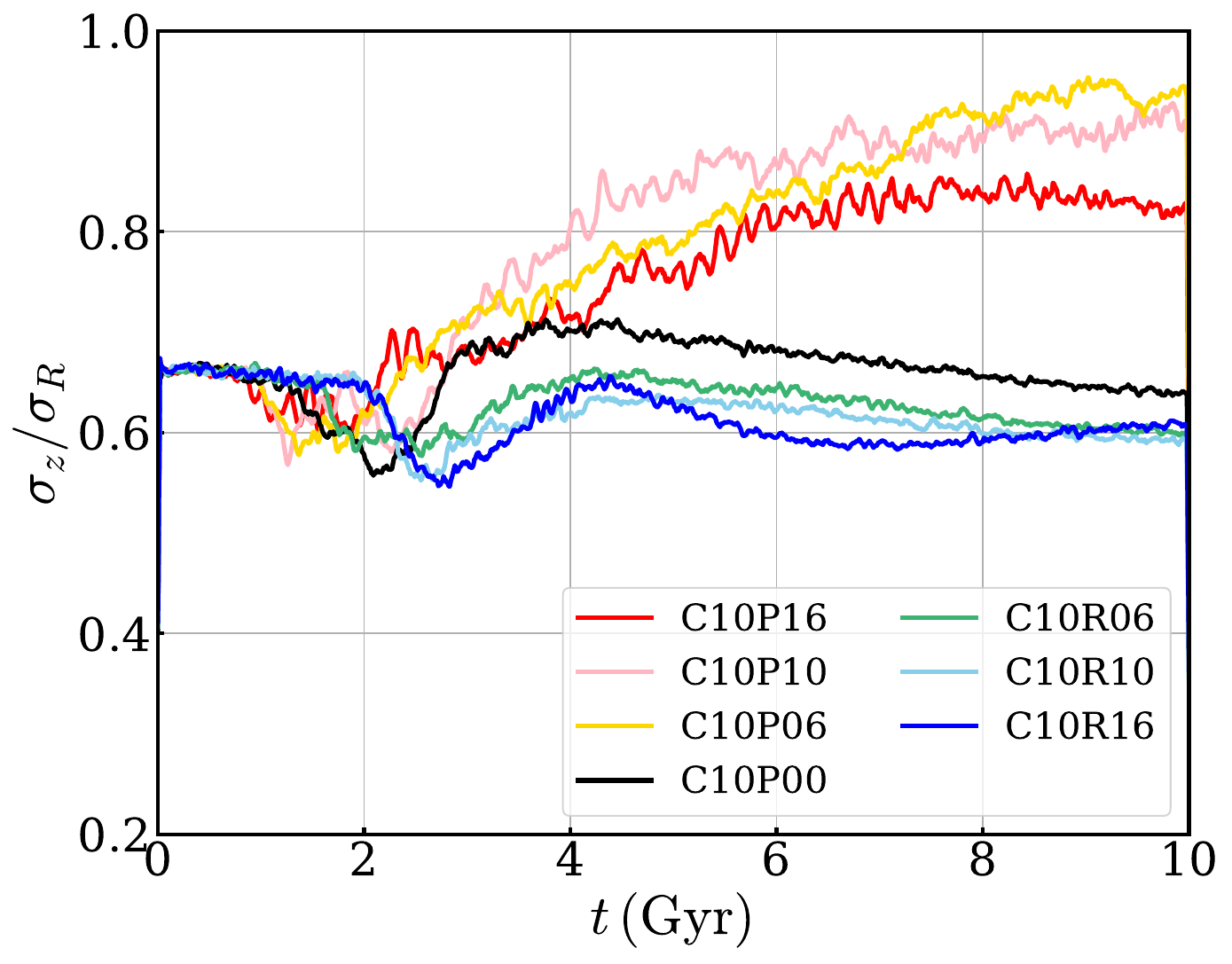}
\caption{Temporal variations of the radio $\sigma_z/\sigma_R$ of the vertical to radial velocity dispersion of the disk particles at $R=2\kpc$ for the models in the \texttt{C00} series (top) and the \texttt{C10} series  (bottom).
\label{fig:sigma}}
\end{figure}

\section{Discussion} \label{sec:discussion}   

We have investigated the formation and evolution of bars in disk galaxies similar to the Milky Way, with a spinning halo.  In this section, we discuss our results compared to the previous investigations. We also use the bar properties to constrain the models for the Milky Way. 

\subsection{Bar Formation under a Spinning Halo} \label{subsec:discuss1} 
In our models, a disk embedded in a faster corotating halo develops a bar earlier, while the bar formation is delayed under a counter-rotating halo, consistent with the results of the previous studies \citep{sn13,long14,col18,col19b}. As mentioned in \autoref{sec:intro}, 
\citet{sn13} tried to explain this in terms of the Ostriker-Peebles parameter $t_{\rm OP}$, arguing that the halo spin increases $t_{\rm OP}$ and thus makes the disk more susceptible to bar formation. Since $t_{\rm OP}$ does not consider the sense of halo rotation, however, it cannot explain why a bar forms faster in a disk under a prograde halo than under the retrograde counterpart.

To account for the disk response to a spinning halo, we examine the growth of various modes in an isolated halo and a disk-halo system. In \autoref{sec:halo}, we present that our spinning halo alone is vulnerable to the growth of $m=2$ spirals due to swing amplification, transforming to a weak bar rotating in the same sense as the halo. 
When a disk is inserted into the halo, $m=2$ spirals in the disk growing also via swing amplification supply gravitational perturbations for the halo. If the halo spin is prograde, the perturbations are well maintained and amplified in the halo, giving positive feedback to the spirals in the disk. Therefore, the disk and the prograde halo work constructively to make the spirals grow faster than in the case with no halo spin, promoting the bar formation (see \autoref{fig:A2A3C00s}). If the halo spin is instead retrograde, $m=2$ perturbations in the disk and halo become out of phase and interact deconstructively. In this case, the $m=2$ spirals in the disk grow more slowly, delaying the bar formation.

\subsection{Buckling Instability} \label{subsec:discuss2}
Numerical simulations commonly found that bars undergo vertical buckling instability when sufficiently strong, resulting in weakened and shorted bars and eventually leading to \ac{BPS} bulges. Our simulations also show that strong bars are subject to buckling instability. There are, however, some discrepancies between the results of our models and the previous simulations \citep{sn13,long14,col18,col19a}. First,  in models of \citet{col18,col19b,col19a}, there is no apparent correlation between $\lambda$ and the epoch of the buckling instability, while the buckling in our models is progressively delayed with increasing $\lambda$. 
Second, the bar strength before the buckling instability is almost independent of $\lambda$ in models with a prograde halo \citep{col18,col19a}
and decreases with $\lambda$ in models with a retrograde halo \citep{col19b}, while it is a decreasing function of $|\lambda|$ for the whole range of $\lambda$ in our models (see \autoref{fig:A2}). Third, the buckling instability almost destroys the bars when $\lambda\gtrsim 0.06$ in \citet{col18,col19a}, while the bars, albeit becoming weaker, still remain strong after the buckling in our models.  Lastly, our models with a bulge do not undergo buckling instability at all, while the models even with a bulge in \citet{li24}  experience buckling instability, although the onset and amplitude of the buckling decrease with increasing bulge mass. 

It is uncertain what causes these differences in the numerical results, but the most likely reason may be differences in the galaxy models adopted. Our models consider a Milky Way-sized galaxy, while the previous authors \citep{long14,col18,col19b,col19a,col21,li24} employed the models with a twice less massive halo and a twice thicker disk than our models. Also, their halo modeled by a truncated NFW profile \citep{nfw} has a different degree of central concentration and a quite different rotation curve from ours (e.g., \autoref{fig:vrot} in the present work can be compared with Figure 1 of \citealt{lieb22}).
These suggest that one should be cautious about drawing a general conclusion by running a limited set of numerical simulations. 

The firehose instability has been invoked as the physical mechanism behind the buckling instability \citep{toomre66,bnt08}.
An alternative mechanism may be the trapping of overlapped planar and vertical 2:1 resonances, proposed recently by \citet{li23}. In the firehose instability, bending perturbations become unstable if the centrifugal force on stars traveling over corrugations exceeds the restoring gravitational force from the other stars, which is achieved when $\sigma_z/\sigma_R$ becomes less than a critical value \citep[see e.g.,][and references therein]{comb81,comb90,raha91,merr94,mar06,kwak17,kwak19,li24}. Our simulations show that the bucking instability in strongly barred galaxies occurs when $\sigma_z/\sigma_R\lesssim 0.47$--$0.60$, roughly consistent with the results of \citet{mar06} and \citet{kwak17}.

The buckling instability in most of our models is consistent with the conventional picture that it needs  $\sigma_z/\sigma_R<0.6$ for operation and that it increases $\sigma_z/\sigma_R$ rapidly, while decreasing $A_2/A_0$, as \cref{fig:A2,fig:sigma} show. However, the buckling instability in model \texttt{C00P16} is exceptional in that it works even with $\sigma_z/\sigma_R\sim0.75$ and does not cause a sharp increase in $\sigma_z/\sigma_R$ and bar weakening. \autoref{fig:vz_z} compares the face-on snapshots of the mean vertical velocities, $\bar v_z$, and the mean vertical positions, $\bar z$, of the disk particles for models \texttt{C00R16} (left) and \texttt{C00P16} (right) each at the epoch of the maximum buckling.  Model \texttt{C00R16} which undergoes strong buckling at $t\sim 2.7\Gyr$ exhibits characteristic quadrupole patterns in the face-on maps, with $|\bar v_z|\lesssim 30\kms$ and 
$|\bar z| \lesssim 0.5\kpc$ \citep[see, e.g.,][]{lokas19,xiang21,li23}. Note that model \texttt{C00P16} also shows similar quadrupole patterns, although they  are weaker with $|\bar v_z| \lesssim 10\kms$ and 
$|\bar z| \lesssim 0.2\kpc$, demonstrating that it indeed suffers the buckling instability near $t\sim9.7\Gyr$. One cannot rule out the possibility of overlapping planar and vertical 2:1 resonances, as envisaged by \citet{li23}, for the nature of the bucking instability in model \texttt{C00P16}.

\begin{figure}[t]
\centering 
\plotone{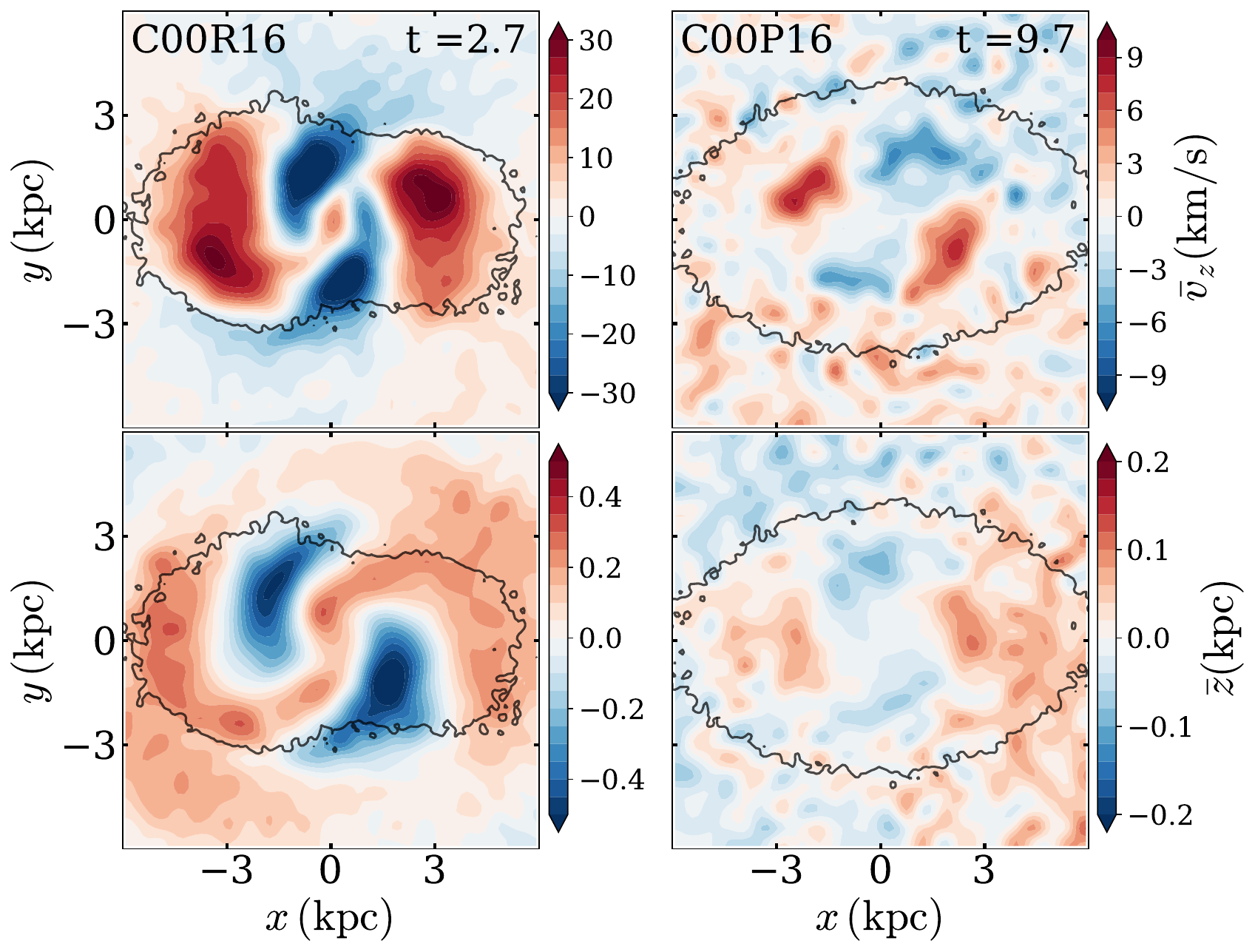} 
\caption{Face-on maps of the mean vertical velocities (top) and mean vertical positions  (bottom) of the stars at $t=2.7\Gyr$ for model \texttt{C00R16} (left) and at $t=9.7\Gyr$ for model \texttt{C00P16} (right). The contour in each panel outlines the bar boundaries defined as the regions with $\Sigma=2\times 10^2\Msun\pc^{-2}$. 
\label{fig:vz_z}}
\end{figure}
 
\subsection{Milky Way Bar} \label{subsec:discuss4}

We now discuss which of our models best describes the properties of the Milky Way bar. Observations show that the Milky Way bar is long with size $R_b \sim 4.5$--$5 \kpc$ and rotates slowly with pattern speed $33 < \Omega_b < 45 \kms \kpc^{-1}$ \citep{weg15,sor15,port17,bg16,cla22}.
Measuring the spin parameter $\lambda$  of a galaxy is a difficult task observationally. 
Recently, \cite{obr22} found a correlation between the angular momenta of dark halos and disks from a sample of galaxies identified in cosmological simulations, and used the relation to estimate $\lambda_{\rm MW}$ of the Milky Way after measuring its disk angular momentum. They obtained $\lambda_{\rm MW} \sim 0.061$ when the dark halo of the Milky Way follows a contracted NFW profile, and $\lambda_{\rm MW} \sim 0.088$ for an uncontracted profile.

\begin{figure}
    \centering
    \plotone{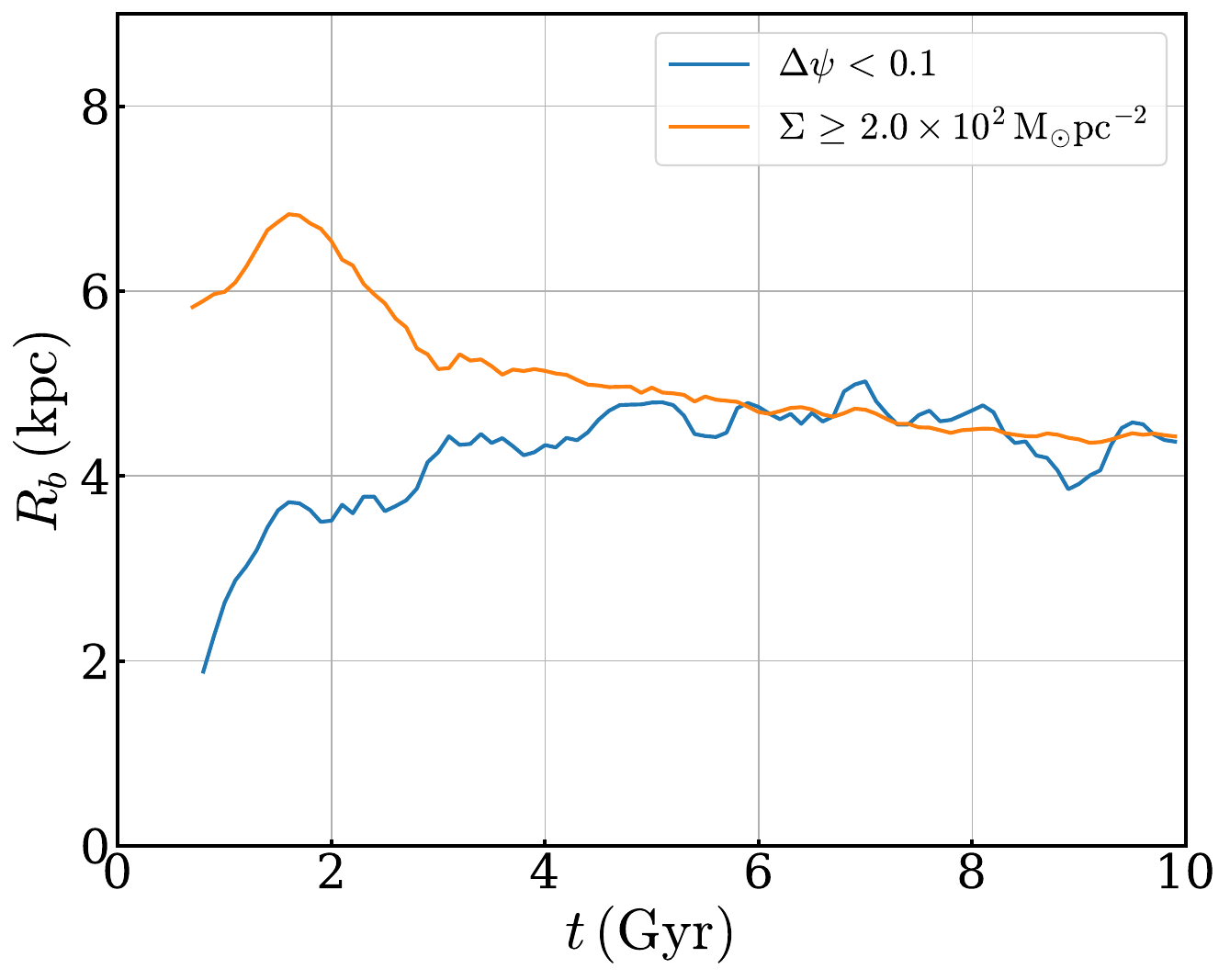}
    \caption{Temporal evolution of the bar length $R_b$ in model \texttt{C10P06}. The blue line corresponds to $R_b$ calculated by imposing the changes of the position angle  $\Delta \psi \leq 0.1$ of the $m=2$ mode, while the orange line is obtained by defining the bar as the regions with $\Sigma\geq2\times10^2\Msun\pc^{-2}$.}
    \label{fig:C10P06}
\end{figure}

\citetalias{jnk23} found that models \texttt{C10} and \texttt{C20} with $\lambda=0$, containing a classical bulge with mass  $10\%$ and $20\%$ of the disk mass, respectively, match the properties of the Milky Way bar quite well. This suggests that model \texttt{C10P06} in the present paper best represents the Milky Way when the halo spin is considered. Indeed, \autoref{fig:omega} shows that the bar in model \texttt{C10P06} has $\Omega_b\sim30$--$35\kms\kpc^{-1}$ for a long period of time.
\autoref{fig:C10P06} plots the temporal changes of the bar length calculated from two methods: (1) the maximum radius where the change in the bar position angle $\Delta \psi<0.1$ and (2) the semi-major axis with $\Sigma \geq 2 \times 10^2 \Msun \pc^{-2}$.  The bar length of $R_b \sim 4$--$5 \kpc$ at $t \gtrsim 3 \Gyr$  is consistent with the observed size of the Milky Way bar. That $R_b$ from $\Delta \psi$ is longer than that from $\Sigma$ at $t\lesssim4\Gyr$ is due to the temporal alignment of the inner parts of the spirals with the bar ends.

We remark on some caveats in concluding that the Milky Way bar is similar to that in model \texttt{C10P06}. 
First, as \cite{obr22} suggested, a contracted NFW profile most likely describes the halo of the Milky Way, while our galaxy models adopt the Hernquist profile.
Second, our models consider only isolated galaxies, while non-axisymmetric perturbations by external triggers such as accretion and minor mergers with satellite galaxies can impact the bar evolution in real situations \citep{Zana18,grand19,pes19, ghosh21,cava22}. The current halo spin of the Milky Way has likely been affected by the most recent major merger. 
Finally, while we adopt gas-free galaxy models, the presence of gas in the disk tends to reduce the bar slow-down rate \citep{seo19,beane23}. To understand the bar evolution of the Milky Way more accurately, it is desirable to run more realistic models by allowing for the halo density distribution, accretion, minor mergers, and the presence of gas and associated star formation.

\section{Conclusions} \label{sec:summary}

We have presented the results of $N$-body simulations to investigate the effects of halo spin and the presence of a classical bulge on the bar formation and evolution. For this, 
we have borrowed two galaxy models from \citetalias{jnk23} that have similar properties to the Milky Way: model \texttt{C00} without a bulge and model \texttt{C10} with a bulge whose mass is 10\% of the disk mass. We have varied the spin parameter $\lambda$ of the halo from $0.16$ (prograde) to $-0.16$ (retrograde) in both series and run the models up to $t = 10 \Gyr$. Our main conclusions are summarized below.

\begin{enumerate}

\item  Simulations of the halo-only models show that our adopted halo is gravitationally stable if it does not rotate ($\lambda=0$). When it rotates with $\lambda=\pm\,0.16$, however, the halo itself is susceptible to forming $m=2$ spirals via swing amplification, eventually growing into a weak bar. The spirals and bar rotate in the same sense as the halo spin, with a pattern speed of $\Omega_h \sim \pm\,7 \kms\kpc^{-1}$. This indicates that our rotating halo \emph{alone} is vulnerable to the formation of non-axisymmetric structures rotating in the same direction as the halo.

\item The tendency of our spinning halo to form $m=2$ non-axisymmetric structures even without external perturbations affects the bar growth time in an embedded disk. When $\lambda>0$, $m=2$ spirals growing via swing amplification in the disk interact constructively with the $m=2$ mode in the prograde halo, promoting the bar formation in the inner disk. When $\lambda$ is large, the spirals in the outer regions are strong enough to limit the further bar growth. When $\lambda<0$, in contrast, the $m=2$ mode in the disk interacts destructively with the $m=2$ mode in the retrograde halo, delaying the bar formation. Since the outer disk is relatively less perturbed by the spirals, the bar in the disk under a counter-rotating halo can grow stronger and longer than that in the models with large $\lambda$.

\item A bar grows by losing its angular momentum to both a halo and bulge, although the amount of angular momentum absorbed by the bulge is much less than that by the halo. The halo particles inside (outside) the corotation resonance $R_\mathrm{CR}$ with the bar can emit (absorb) angular momentum to (from) the bar. Since $R_\mathrm{CR}$ is larger for larger $\lambda$, the angular momentum transfer from disk to halo tends to be larger, causing the bar pattern speed $\Omega_b$ to decay faster, for smaller $\lambda>0$.  In models with $\lambda=0.16$,  the angular momentum absorption and emission are almost balanced, keeping $\Omega_b$ and bar strength nearly constant for a long period. Under a retrograde or non-spinning halo, in contrast, all the halo particles lag behind the bar and thus can absorb angular momentum from the bar efficiently. This causes $\Omega_b$ to decay relatively fast, almost independent of $\lambda\leq0$.

\item All our models form a bar, and models with a strong bar develop a \ac{BPS} bulge. In our models, the presence of a classical bulge reduces the bar growth and suppresses the buckling instability completely. In all models without a classical bulge (\texttt{C00} series), a \ac{BPS} bulge is produced rapidly through the buckling instability. In models with a classical bulge (\texttt{C10} series), however, bars slowly thicken vertically via gravitational interactions with the disk particles.

\item In models with no classical bulge, the buckling instability tends to occur earlier along the sequence with smaller $\lambda$: a bar under a retrograde halo deforms earlier in the vertical direction than the prograde counterpart. For most of our models, the onset of the buckling instability corresponds to the ratio of the velocity dispersion $\sigma_z/\sigma_R\lesssim0.47$--$0.60$ at $R=2\kpc$, roughly consistent with the previous results \citep[e.g.,][]{mar06,kwak17}, except model \texttt{C00P16} which undergoes mild buckling instability even at $\sigma_z/\sigma_R \sim0.75$. The models with a bulge have $\sigma_z/\sigma_R\gtrsim 0.55$ and remain stable to buckling instability.

\item Among our models, model \texttt{C10P06} is similar to the Milky Way in terms of the halo spin and bar properties. It has the halo spin parameter of $\lambda=0.06$, identical to the observationally inferred value \citep{obr22}, and a classical bulge with mass 10\% of the disk mass. The bar in this model has a semi-major axis of $R_b \sim 4$--$5 \kpc$ and a pattern speed of $\Omega_b\sim30$--$35\kms\kpc^{-1}$ for an extended period of time, which are consistent with the observed properties of the Milky Way bar.

\end{enumerate}

\section*{acknowledgments}
The work of D.J.\ was supported by Basic Science Research Program through the National Research Foundation of Korea (NRF) funded by the Ministry of Education (RS-2023-00273275). The work of W.-T.\ K.\ was supported by a grant of the National Research Foundation of Korea (2022R1A2C1004810).  Computational resources for this project were provided by the Supercomputing Center/Korea Institute of Science and Technology Information with supercomputing resources including technical support (KSC-2023-CRE-0175).


\end{document}